\documentclass{aa}
\usepackage[bookmarks,bookmarksopen,colorlinks,linkcolor={blue},citecolor={blue},
  urlcolor={red}]{hyperref}
\usepackage{amsmath}
\usepackage{natbib}
\usepackage{color}
\usepackage[x11names]{xcolor}
\usepackage{graphics}
\usepackage{geometry}
\usepackage{enumerate}
\usepackage{setspace}
\usepackage{siunitx}
\usepackage{pdflscape}
\newcommand{\prot}{$P_{\text{rot}}$}
\newcommand{\mps}[1]{m s$^{-1}$#1}
\defcitealias{benneke17}{B17}

\newcommand{\cdbox}[1]{%
  \colorlet{currentcolor}{.}%
  {\color{Blue1}%
    \dbox{\color{currentcolor}#1}}%
}
\usepackage{dashbox,framed,color,ocg-p}
\newcommand{\ToggleLayer}[2]{%
  \leavevmode
  \pdfstartlink user {
    /Subtype /Link
    /Border [0 0 0]%
    /A <<
      /S/JavaScript
      /JS (
         var aOCGs = this.getOCGs(), Layer;
         var Layers = "#1".split(","), Active = -1, i, l;
         for (l=0; l<Layers.length; l++) {
           Layer = Layers[l];
           for (i=0; aOCGs && i<aOCGs.length; i++) {
             if (aOCGs[i].state && aOCGs[i].name == Layer) {
               Active = l;
               aOCGs[i].state = false;
             }
           }
           if (Active >= 0) break;
         }
         if (Active == -1) {
           for (l=0; l<Layers.length; l++) {
             if (Layers[l] == "") Active = l;
           }
         }
         Active = Active + 1;
         if (Active == Layers.length) Active = 0;
         Layer = Layers[Active];
         for (i=0; aOCGs && i<aOCGs.length; i++) {
           if (aOCGs[i].name == Layer) aOCGs[i].state = true;
         }
      )
    >>
  }#2%
  \pdfendlink
}

\begin{document}

\title{Characterization of the K2-18 multi-planetary system with HARPS}
\subtitle{A habitable zone super-Earth and discovery of a second, warm super-Earth on a non-coplanar orbit}
\titlerunning{Two super-Earths around K2-18}
\authorrunning{Cloutier et al.:}

\author{R.~Cloutier \inst{1,2,3}
  \and N.~Astudillo-Defru \inst{4}
  \and R.~Doyon \inst{3}
  \and X.~Bonfils \inst{5}
  \and J.-M.~Almenara \inst{4}
  \and B.~Benneke \inst{3}
  \and F.~Bouchy \inst{4}
  \and X.~Delfosse \inst{5}
  \and D.~Ehrenreich \inst{4}
  \and T.~Forveille \inst{5}
  \and C.~Lovis \inst{4}
  \and M.~Mayor \inst{4}
  \and K.~Menou \inst{1,2}
  \and F.~Murgas \inst{5}
  \and F.~Pepe \inst{4}
  \and J.~Rowe \inst{3}
  \and N.~C.~Santos \inst{6,7}
  \and S.~Udry \inst{4}
  \and A.~W\"unsche \inst{5}
}

\institute{Dept. of Astronomy \& Astrophysics, University of Toronto, 50 St. George Street, M5S 3H4, Toronto, ON, Canada
  \email{cloutier@astro.utoronto.ca}
  \and Centre for Planetary Sciences, Dept. of Physical \& Environmental Sciences, University of Toronto Scarborough, 1265 Military Trail, M1C 1A4, Toronto, ON, Canada
  \and Institut de Recherche sur les Exoplan\`etes, d\'epartement de physique, Universit\'e de Montr\'eal, C.P. 6128 Succ. Centre-ville, H3C 3J7, Montr\'eal, QC, Canada
  \and Observatoire Astronomique de l’Universit\'e de Gen\`eve, 51 chemin des Maillettes, 1290 Versoix, Switzerland
  \and Universit\'e Grenoble Alpes, CNRS, IPAG, F-38000 Grenoble, France
  \and Instituto de Astrof\'isica e Ci\^encias do Espa\c{c}o, Universidade do Porto, CAUP, Rua das Estrelas, 4150-762 Porto, Portugal
  \and Departamento de F\'isica e Astronomia, Faculdade de Ci\^encias, Universidade do Porto, Rua do Campo Alegre, 4169-007 Porto, Portugal
}

\abstract{}{The bright M2.5 dwarf K2-18 ($M_s=0.36$ M$_{\odot}$, $R_s=0.41$ R$_{\odot}$)
  at 34 pc is known to host a transiting super-Earth-sized
  planet orbiting within the star's habitable zone; K2-18b. Given the superlative nature
  of this system for studying an exoplanetary atmosphere receiving similar levels of insolation
  as the Earth, we aim to characterize the planet's mass which is required to interpret
  atmospheric properties and infer the planet's bulk composition.}{We obtain precision radial velocity
  measurements with the HARPS spectrograph and couple those measurements with the \emph{K2} photometry
  to jointly model the observed
  radial velocity variation with planetary signals and a radial velocity jitter
  model based on Gaussian process regression.}{We measure the mass of K2-18b to be $8.0 \pm 1.9$
  M$_{\oplus}$ with a bulk density of $3.7 \pm 0.9$ g/cm$^3$ which may correspond to a predominantly
  rocky planet with a significant gaseous envelope or an ocean planet with a water mass fraction $\gtrsim 50$\%.
  We also find strong evidence for a second, warm super-Earth K2-18c ($m_{p,c}\sin{i_c} = 7.5 \pm 1.3$ M$_{\oplus}$)
  at $\sim 9$ days with a semi-major axis $\sim 2.4$ times smaller than the transiting K2-18b.
  After re-analyzing the available light
  curves of K2-18 we conclude that K2-18c is not detected in transit and therefore likely has an orbit
  that is non-coplanar with the orbit of K2-18b. A suite of dynamical integrations are performed
  to numerically confirm the system's dynamical stability. By varying the simulated orbital
  eccentricities of the two planets, dynamical stability constraints are used
  as an additional prior on each planet's eccentricity posterior from which we
  constrain $e_b < 0.43$ and $e_c < 0.47$ at the level of 99\% confidence.}{The discovery of the inner
  planet K2-18c further emphasizes the prevalence of multi-planet systems around M dwarfs. The
  characterization of the density of K2-18b reveals that the planet likely has a thick gaseous envelope
  which along with its proximity to the Solar system makes the K2-18 planetary system an interesting target
  for the atmospheric study of an exoplanet receiving Earth-like insolation.}

\maketitle

\section{Introduction}

Exoplanets orbiting within their host star's habitable zone may have surface temperatures conducive
to allowing for the presence of liquid water on their surfaces depending on the properties of the planetary
atmosphere \citep{kasting93}; a condition likely required to sustain extraterrestrial life.
This implies that habitable zone exoplanets receive comparable levels of stellar insolation as the Earth
does from the Sun. Habitable zone exoplanets therefore represent superlative opportunities to
search for life outside of the Solar system via the characterization of their atmospheric structure
and composition via transmission spectroscopy for transiting exoplanets.


M dwarf host stars are ideal targets to probe potentially habitable exoplanetary atmospheres
\citep[e.g.][]{kaltenegger11, rodler14}. 
Transmission spectroscopy observations of transiting habitable zone (HZ) exoplanets around
M dwarfs are favourable compared
to around Sun-like stars given the increased depth of the transit for a given sized planet  
\citep[e.g.][]{stevenson10, kreidberg14a}. In addition, the orbital periods corresponding to
the HZ are less than around Sun-like stars (weeks compared to 12 months)
thus increasing the number of accessible transit events within a given observational baseline.
M dwarfs are also known to frequently host multiple small planets
\citep[typically 2.5 planets per star with $0.5 \leq r_p/\text{R}_{\oplus} \leq 4$ and within 200 days;][]{dressing15a,gaidos16}
thus enabling direct comparative planetology to be conducted on known multi-planet systems.

\cite{montet15} reported the detection of the HZ super-Earth K2-18b originally proposed in
the \emph{K2} light curve analysis of \cite{foremanmackey15b}. In these studies,
two transit events were observed in Campaign 1 data from the
re-purposed \emph{Kepler} spacecraft mission \emph{K2} whose field coverage only lasted for
$\sim 80$ days. The existence of the planet was
confirmed and uncertainties regarding its ephemeris were significantly reduced in
\cite{benneke17} (hereafter \citetalias{benneke17})
who used follow-up transit observations with the \emph{Spitzer Space Telescope} to
detect an additional transit event. The now confirmed
super-Earth K2-18b orbits an M2.5 dwarf with a period of $\sim 32.9$ days placing it directly within
the star's habitable zone \citep{kopparapu13}. The measured radius of
2.28 R$_{\oplus}$ is suggestive of an extended H/He envelope \citep{valencia13, rogers15, fulton17} that may
contain additional molecular species such as water and/or methane that could be detectable with the
\emph{James Webb Space Telescope} \citep[JWST;][]{beichman14}.
Owing to the proximity of the system
\citep[$\sim 34$ pc, $V=13.5$, $I=11.7$, $K=8.9$;][]{cutri03, zacharias13}, K2-18
is truly an attractive target for characterizing the atmosphere of a HZ super-Earth with
unprecedented precision in the JWST-era.

In this study we report the first measurement of the planetary mass of K2-18b using precision
radial velocity measurements taken with the HARPS spectrograph. In this data we also find strong
evidence for an additional super-Earth whose orbit is interior to K2-18b but is not found
to transit. In Sect.~\ref{sect:obs} we summarize the HARPS spectroscopic and K2 photometric
observations used in our analysis,
in Sect.~\ref{sect:periodograms} we analyze the periodic signals in the spectroscopic data
and in Sect.~\ref{sect:joint} we discuss our various radial velocity modelling procedures.
In Sect.~\ref{sect:results} 
we present the results of our radial velocity analysis including the detection of a second 
super-Earth K2-18c in the system which we show is non-transiting and therefore non-coplanar with
K2-18b in Sect.~\ref{sect:transit}.
Lastly we perform a dynamical analysis of the two-planet system in Sect.~\ref{sect:dynam} to
dynamically constrain the orbital eccentricities of the planets  
before concluding with a discussion in Sect.~\ref{sect:disc}.

\section{Observations} \label{sect:obs}
\subsection{HARPS Spectra}
From April 2015 (BJD=2457117.5) to May 2017 (BJD=2457875.5),
we collected 75 spectra of K2-18 (EPIC 201912552) with the high-resolution (R=115000)
HARPS spectrograph \citep{mayor03,pepe04}. The majority of exposure
times were fixed to 1800 seconds with the exception of the following six epochs
whose exposure times were modified to the following reported values: 2400 seconds
(BJD-2,450,000 = 7199.503915, 7200.503114), 1200 seconds (BJD-2,450,000 = 7204.491167),
and 900 seconds (BJD-2,450,000 = 7810.806284, 7814.760772, 7815.759421). 
The online HARPS pipeline returns the extracted, wavelength-calibrated
spectra \citep{lovis07}. Initial radial velocity estimates are computed
from the cross-correlation of each spectrum with 
a numerical mask \citep{baranne96, pepe02}.
Using each spectrum's initial estimate, all spectra are shifted to
a common reference frame by their corresponding barycentric correction such
that spectral features originating from the target star become aligned while
telluric features are shifted by minus the epoch's barycentric correction.
The median combination of these shifted spectra is then used to construct
a custom reference spectrum at high signal-to-noise (S/N). A telluric template is
then constructed from the median combination of all residual spectra after
removal of the high S/N reference stellar spectrum. The process of computing the
median reference stellar spectrum is then repeated using the individual spectra
with tellurics masked by the median telluric spectrum. 
We then compute precision radial velocities by performing a $\chi^2$-minimization
of each spectrum with the reference spectrum \citep{astudillodefru15}.
Radial velocity uncertainties are then estimated directly on the reference
spectrum \citep{bouchy01}.

From the extracted spectra we also derive a number of activity indicators including
the time-series of the H$\alpha$ index which is sensitive to chromospheric activity and
is computed following the definition in \cite{bonfils07}. For the M dwarf K2-18
\citep[$V$=13.5;][]{henden14} the H$\alpha$ index is favoured over the \ion {Ca} {II}~H+K
Mt. Wilson S index \citep{wilson68, baliunas95}
due to the low S/N obtained in the blue. From the S index we
derive $\log{R'_{\text{HK}}}=-5.247 \pm 0.318$ \citep{astudillodefru17a}. Additionally we derive
the full width at half maximum (FWHM) and bi-sector inverse slope (BIS)
shape parameters of the cross-correlation function which are modified
by dark and/or bright active regions traversing the visible stellar surface. In
Sect.~\ref{sect:periodograms} we'll use these ancillary time-series to learn about
the star's activity simultaneously with our radial velocity measurements. 
All spectroscopic time-series are reported in Table~\ref{table:data}.

\subsection{K2 Photometry}
K2-18 was observed in long-cadence mode during Campaign 1 of the
K2 mission as part of the `Targeting M dwarfs with K2' proposal
(GO1053\footnote{K2-18 was also targeted in the following programs:
  GO1006, GO1036, GO1050, GO1051, GO1052, GO1059, GO1063, GO1075}, PI: B. Montet).
The baseline of the K2 light curve is just 80 days but provides nearly continuous
coverage between June 1st, 2014 (BJD=2456810.5) and August 20th, 2014 (BJD=2456890.5). 

We obtained the full de-trended light curve from the
MAST\footnote{http://archive.stsci.edu/k2/hlsp/everest/search.php} data retrieval service. As a result
of the loss of two reaction wheels on-board the Kepler spacecraft, photometric
observations from the K2 mission exhibit a reduced pointing precision, and
hence photometric precision, compared to the original Kepler mission. Raw K2
light curves must be de-trended with the variable pointing of the
spacecraft throughout the observing sequence. We select the \emph{EVEREST} reduction of
the K2 light curve which performs this de-trending correction \citep{luger16}.

\begin{figure*}
  \centering
  \includegraphics[width=\hsize]{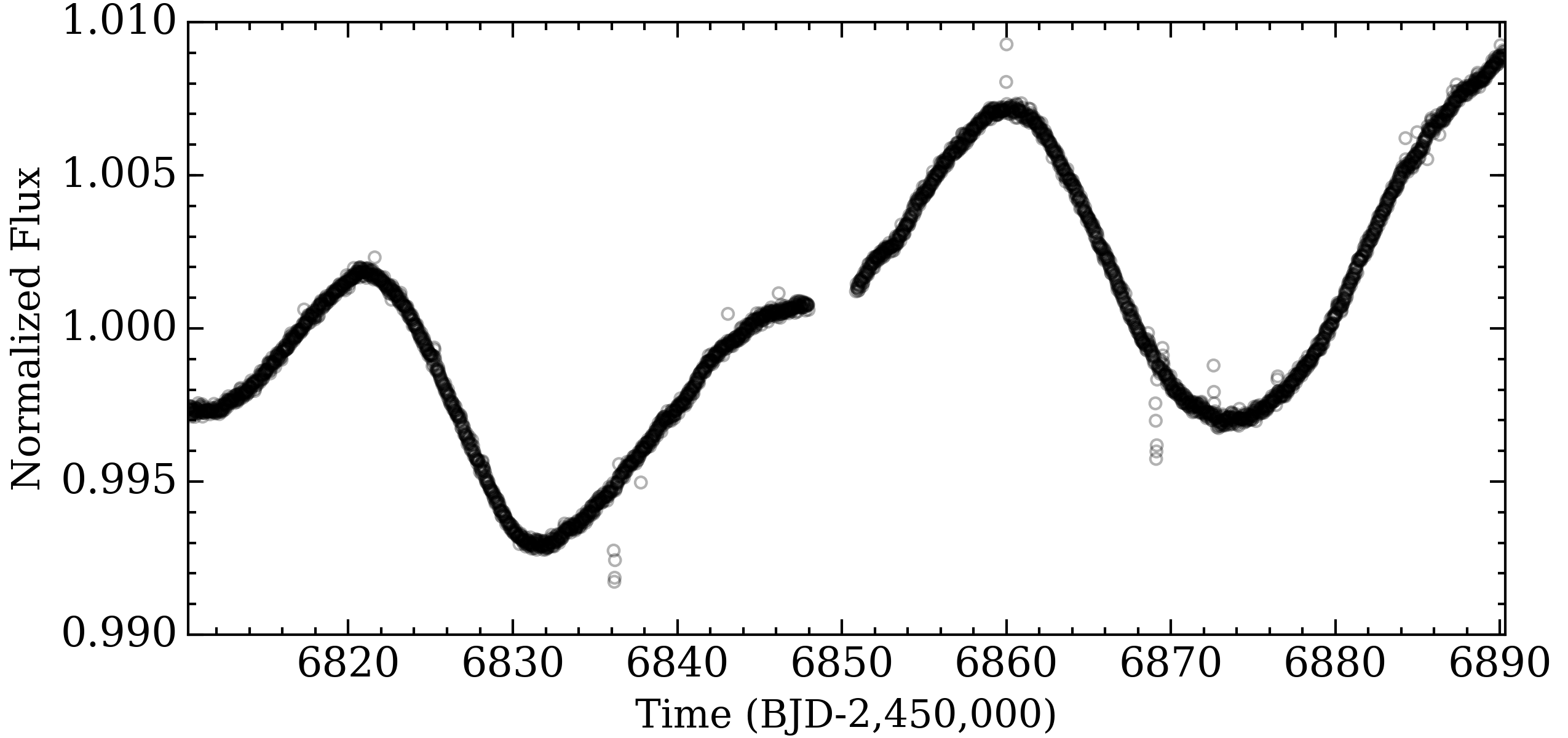}%
   \hspace{-\hsize}%
  \begin{ocg}{fig:ticksoff}{fig:ticksoff}{0}%
  \end{ocg}%
  \begin{ocg}{fig:tickson}{fig:tickson}{1}%
    \includegraphics[width=\hsize]{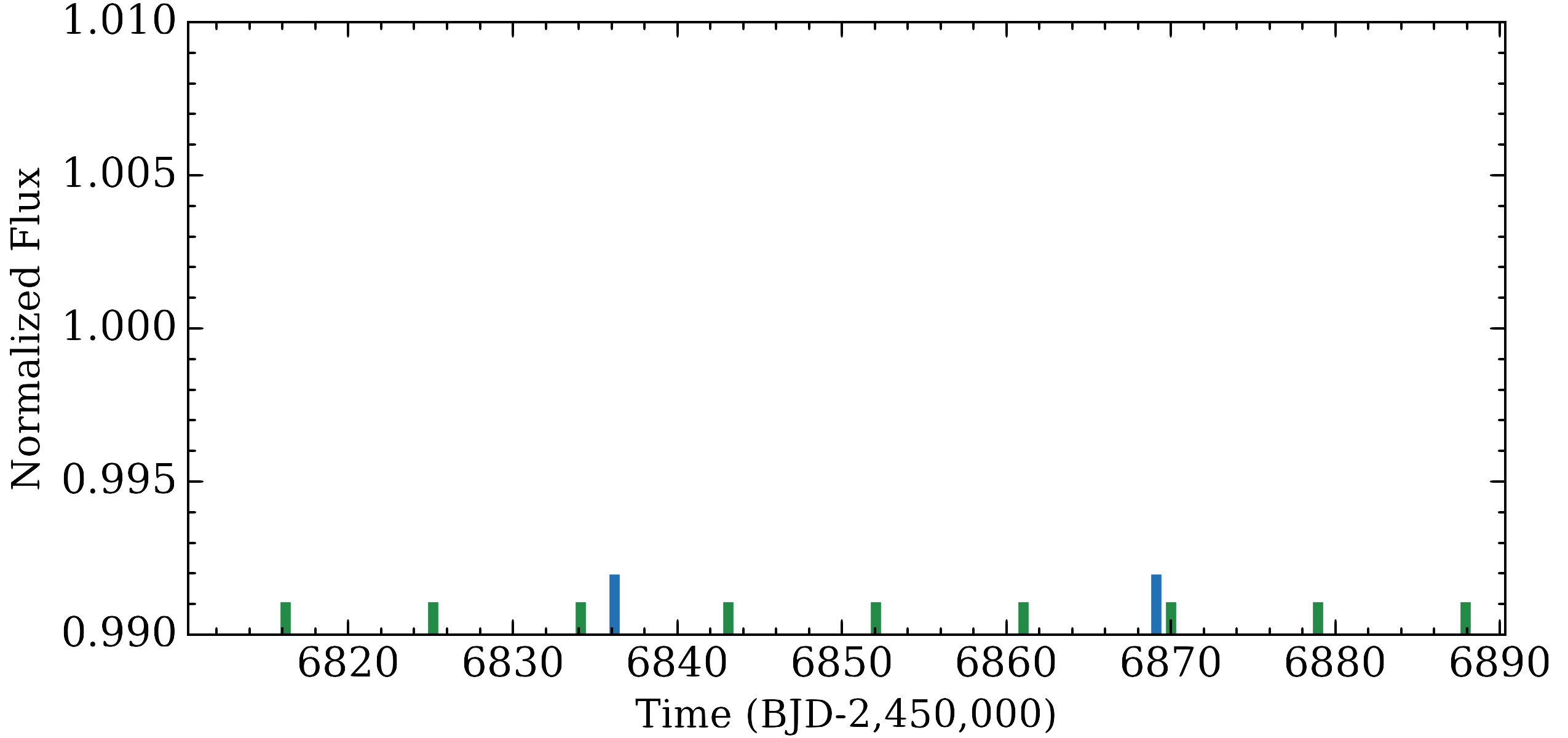}%
  \end{ocg}
  \hspace{-\hsize}%
  \begin{ocg}{fig:gpoff}{fig:gpoff}{0}%
  \end{ocg}%
  \begin{ocg}{fig:gpon}{fig:gpon}{1}%
    \includegraphics[width=\hsize]{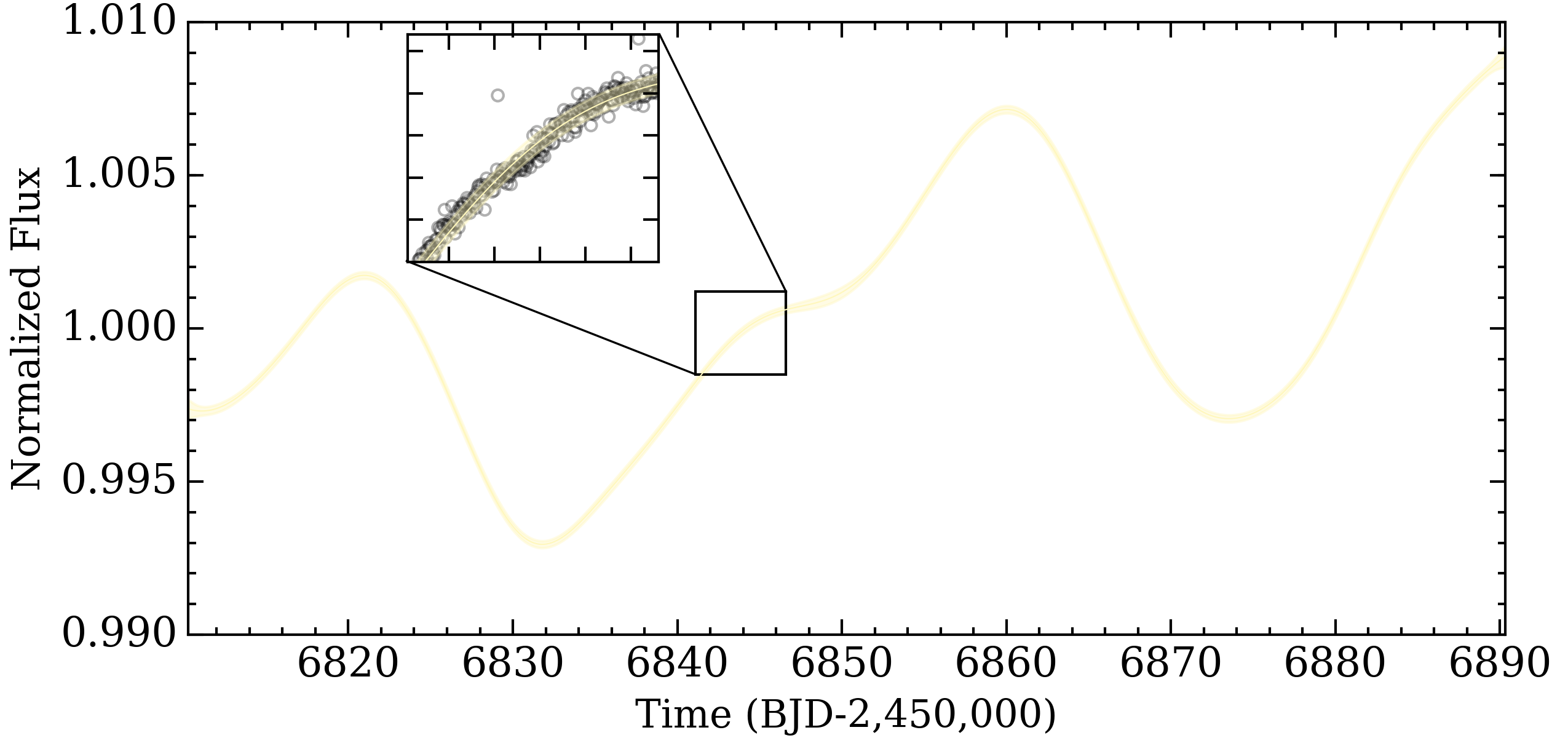}%
  \end{ocg}%
  \caption{The K2 photometric light curve after the removal of known unphysical spurious signals.
    The two \ToggleLayer{fig:tickson,fig:ticksoff}{\protect\cdbox{transits}} of K2-18b are highlighted by the
    \emph{long blue ticks} with the expected times of
    mid-transit for K2-18c highlighted with \emph{short green ticks} (see Sect.~\ref{sect:results}).
    The \ToggleLayer{fig:gpon,fig:gpoff}{\protect\cdbox{\emph{solid yellow curve}}} is the mean of the predictive
    GP distribution and the surrounding shaded regions mark its 99\% confidence intervals. The
    upper left sub-panel is a magnified view of the highlighted region to aid in the visualization of the
    data and the GP fit.}
  \label{fig:k2phot}
\end{figure*}

The majority of the residual photometric variability following de-trending of
the light curve can be attributed to the intrinsic photometric variability of
the star and two observed transits of K2-18b from \cite{montet15}. Removal
of the transit events provides a dataset that can be used to investigate the
photometric stellar jitter resulting from active regions
traversing the visible stellar surface thus giving rise to the star's observed
photometric variability. For reference, the de-trended light curve
is shown in Fig.~\ref{fig:k2phot} along with our Gaussian process fit to the
light curve (see Sect.~\ref{sect:gp} for an explanation of the fit).

\section{Periodogram Analysis} \label{sect:periodograms}
\begin{figure}
  \centering
  \includegraphics[width=\hsize]{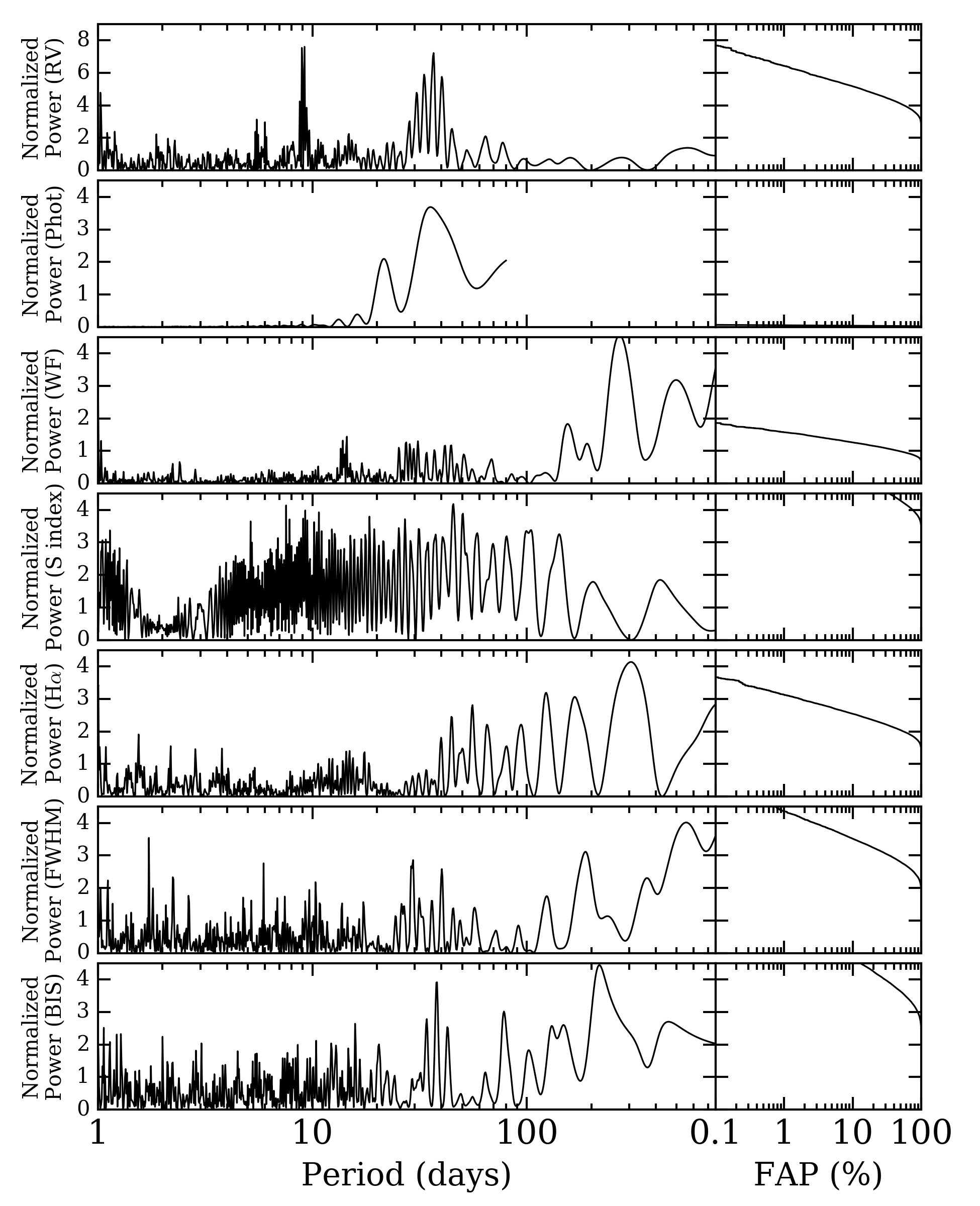}%
   \hspace{-\hsize}%
  \begin{ocg}{fig:Psoff}{fig:Psoff}{0}%
  \end{ocg}%
  \begin{ocg}{fig:Pson}{fig:Pson}{1}%
    \includegraphics[width=\hsize]{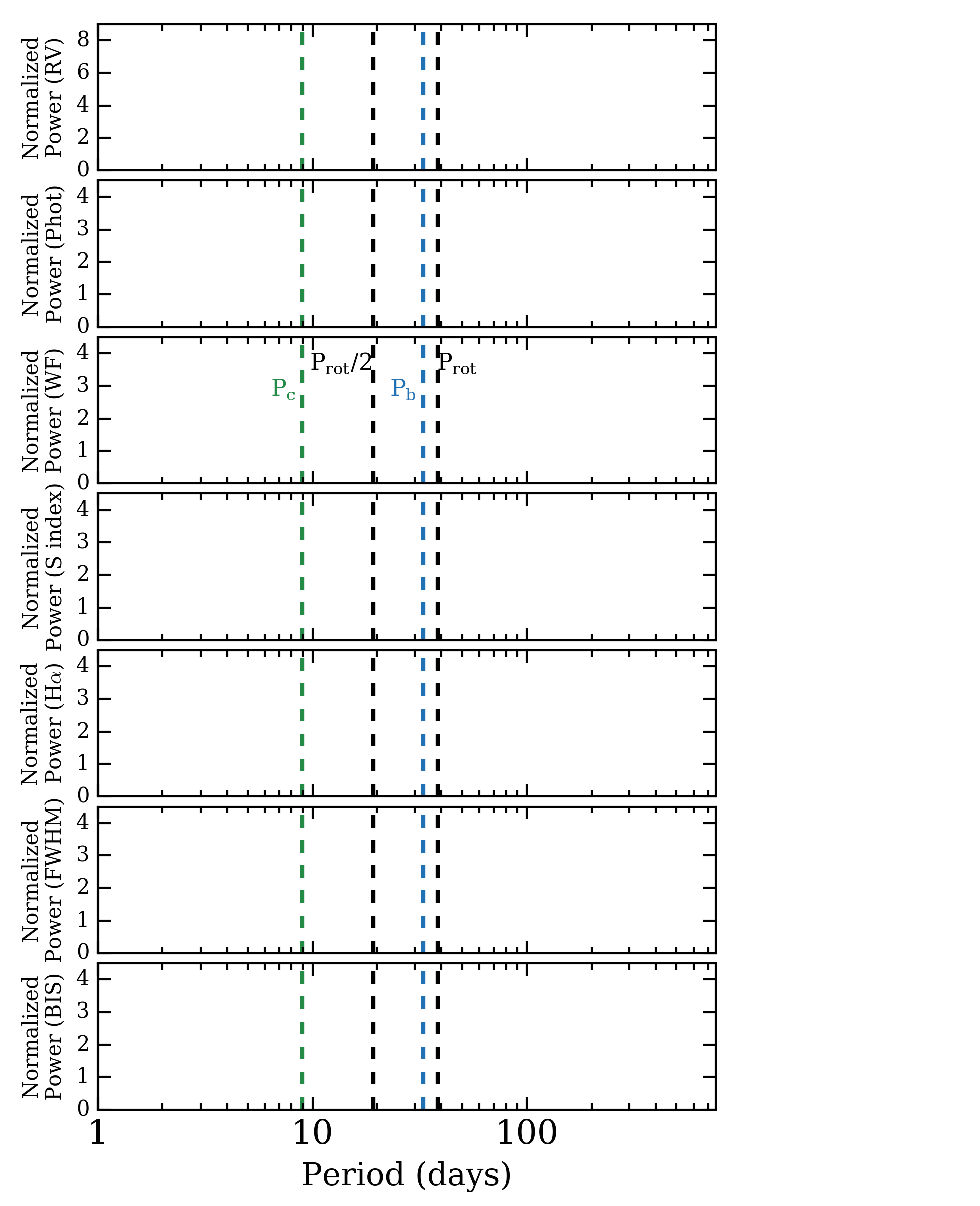}%
  \end{ocg}
  \hspace{-\hsize}%
  \caption{\emph{Left column (top to bottom)}:
  Lomb-Scargle periodograms of the raw radial velocities (RV), the K2 photometry (Phot), the HARPS window
  function (WF), S index, H$\alpha$, full width half maximum (FWHM), and bi-sector inverse slope (BIS)
  time-series. The orbital periods of K2-18b, K2-18c, the stellar rotation
  period, and its first harmonic are highlighted with
  \ToggleLayer{fig:Pson,fig:Psoff}{\protect\cdbox{\emph{vertical dashed lines}}}.
  \emph{Right column}: the false alarm probability as a function of normalized periodogram power for each
  time-series. The FAP curve for the photometry spans very low power and is barely discernible in its
  subpanel. Suffice it to say that any signal with normalized power $> 10^{-2}$ has a FAP $\ll 0.1$\%.
  The FAP curve for the S index exhibits FAP $\gtrsim 30$\% for all power visible on its ordinate and 
  therefore is only visible in the upper right of its subpanel.}
  \label{fig:periodograms}
\end{figure}

Accurate modelling of the stellar radial velocity (RV) variations requires knowledge of
the strong periodicities present in the data. These signals include contributions from
both orbiting
planets and from the rotation of active regions present on the stellar surface which
give rise to stellar RV jitter and are modulated by the stellar rotation period
and/or its harmonics. In the \emph{top panel} of Fig.~\ref{fig:periodograms} we plot the
Lomb-Scargle periodogram \citep{scargle82} of the raw RVs to determine which periodicities are
present at high significance i.e. with a low false alarm probability (FAP). In all
LS-periodograms we calculate FAPs via bootstrapping with replacement using $10^4$ iterations
and individually normalize each periodogram's power by its standard deviation. 

Two important features are detected in the LS-periodogram of the RVs.
The first is a forest of peaks ranging from 
$\sim 25-45$ days with distinct peaks centred on both the orbital period of K2-18b
($P_b \sim 33$ days \citetalias{benneke17}; FAP = 2.9\%) and the approximate stellar
rotation period from the K2 photometry (see Sect.~\ref{sect:gp} for fitting of the
stellar rotation period \prot{} $\sim 38.6$ days;
FAP = 0.1\%). The second important feature is a pair of closely spaced peaks
at $\sim 9$ days (FAP $<0.01$\%) which because of their similar period and power
likely result from a single source. This feature at $\sim 9$ days
constitutes the strongest periodic signal in the
periodogram of the raw RVs and is not observed in the periodograms of any of the ancillary
time-series, nor in the periodogram of the window function, all of which are
shown in the remaining panels of Fig.~\ref{fig:periodograms}. The aforementioned time-series
include the K2 photometry (see Fig.~\ref{fig:k2phot}), the window function or
time sampling of the HARPS observations, and four spectroscopic activity indicators:
the S index, H$\alpha$ index, FWHM, and the BIS of the cross-correlation function. 
Together the presence of the strong $\sim 9$ day signal in radial velocity
and its absence elsewhere provides strong initial evidence for a second planet in the K2-18
system at $\sim 9$ days.

\section{Joint Modelling of Planets and RV Jitter} \label{sect:joint}
\subsection{Training the GP jitter model on ancillary time-series} \label{sect:gp}
The K2 photometry of K2-18 exhibits quasi-periodic photometric variability with a
semi-amplitude of $\sim 0.008$ mag and a rotation period of \prot{} $\sim 38.6$ days as
seen in Fig.~\ref{fig:k2phot}. This makes K2-18
a moderately active early M dwarf in terms of its photometric variability \citep{newton16a}
the origin of which likely results from the rotation of active regions across the projected stellar
disk at or close to \prot{} owing to the characteristically low amplitudes of differential
rotation in M dwarfs \citep{kitchatinov11}. 
The observed photometric variability---or photometric jitter---has
a correlated manifestation in the
variation of the star's apparent radial velocity and certain spectroscopic indicators
because it is a single physical process that is responsible for the jitter in each
time-series \citep{aigrain12}.

In order to obtain accurate and self-consistent detections of the planetary signals in radial velocity
we must model the RV jitter of K2-18 simultaneously with our planet model. Because the stellar 
photometric rotation period of $\sim 38$ days (see second panel in Fig.~\ref{fig:periodograms}) is
marginally detected in the LS-periodogram of the RVs and unambiguously detected in the K2
photometry, we consider in our first mode---called Model 1---the
K2 photometric light curve, less the observed transits of K2-18b
to train our RV jitter model whose covariance
properties are common with the photometric variability. However two important caveats exist when adopting
the K2 photometry as our training set. The first being that the baseline of the photometry spans just 80 days
implying that any temporal variation whose characteristic timescale is greater than this baseline will
remain unconstrained or at best weakly constrained. Secondly the K2 photometry were obtained nearly 8 
months prior to our HARPS observations such that any evolution in the jitter's covariance structure between
observing sequences from say magnetic activity cycles, will not be captured in the training set. For these
reasons we will also consider the BIS time-series as an alternative training set in a
second round of modelling called Model 2.
Being contemporaneous with the RV measurements, training on the BIS time-series mitigates the two
aforementioned issues. In place of the BIS we also tested training on the S index, H$\alpha$, and
FWHM time-series but find results consistent with training on the BIS. Following \cite{faria16} we also
consider a joint jitter + planet model but neglect any training of the jitter model's covariance structure in
a third model; Model 3. Finally for comparison
purposes we will also consider a fourth model---called Model 4---that neglects any contribution from stellar
jitter.

To implement this joint modelling procedure
we follow \cite{cloutier17} by using a Gaussian process (GP) regression model to model the covariance between
adjacent observations in our training sets where applicable (i.e. in Models 1 and 2).
GP regression is an attractive method for modelling the
stochastic processes that give rise to observable RV jitter as it is \emph{non-parametric} and therefore
independent of an assumed functional form of the jitter. The GP prior is represented by a multi-variate
Gaussian distribution of functions described by a covariance matrix
$K_{ij}=k_{ij}(\boldsymbol{\theta}) + \sigma^2_i\delta_{ij}$
with a function $k_{ij}(\boldsymbol{\theta})=k(t_i,t_j,\boldsymbol{\theta})$ that parameterizes the
covariance between values of the observable $\mathbf{y}(\mathbf{t})$ at the epochs $t_i$ and $t_j$ in
$\mathbf{t}$. The observable $\mathbf{y}(\mathbf{t})$ has associated uncertainties
$\boldsymbol{\sigma}(\mathbf{t})$
which are added along the diagonal of the covariance matrix $K$ in quadrature. 
The set of GP hyperparameters $\boldsymbol{\theta}$ are unique to the chosen covariance function
$k_{ij}(\boldsymbol{\theta})$ and are solved for in the training step. After solving for the
GP hyperparameters and thus obtaining a unique GP prior distribution, the GP prior conditioned on the data
$\mathbf{y}(\mathbf{t})$ becomes the predictive distribution. The mean function of the GP predictive distribution
can be evaluated at previously unseen epochs $\mathbf{t^*}$ using 

\begin{equation}
  \boldsymbol{\mu}(\mathbf{t^*}) = K(\mathbf{t^*},\mathbf{t}) \cdot K(\mathbf{t},\mathbf{t})^{-1} \cdot
  \mathbf{y}(\mathbf{t}), \label{eq:meanGP}
\end{equation}

\noindent which we take to be our GP jitter model of the RVs 
by evaluating Eq.~\ref{eq:meanGP} at $\mathbf{t}$.

Because the stellar jitter, and in particular the photometric jitter,
is modulated by the stellar rotation period, we include a
periodic term in our assumed covariance function $k_{ij}(\boldsymbol{\theta})$
with period equal to \prot{.} We also include a radial component 
due to the stochastic temporal evolution of starspot lifetimes, spatial distributions,
and contrasts thus forcing the covariance to not be strictly periodic.
Explicitly the adopted covariance structure is
parameterized by a quasi-periodic covariance kernel of the form

\begin{equation}
k_{i,j}(\boldsymbol{\theta}) = a^2 \exp{\left[ - \frac{|t_i-t_j|^2}{2\lambda^2} -\Gamma^2
    \sin^2{\left(\frac{\pi|t_i-t_j|}{P_{\text{GP}}} \right)} \right]},
\label{eq:cov}
\end{equation}

\noindent which is parameterized by four hyperparameters
$\boldsymbol{\theta}=(a,\lambda,\Gamma,P_{\text{GP}})$: $a$ the amplitude of the
correlations, $\lambda$ the exponential timescale, $\Gamma$ the coherence scale of the
correlations, and $P_{\text{GP}}$ the periodic timescale of the correlations which we
interpret as \prot{.} We also include an additional
scalar jitter parameter $s$ which is added in quadrature to the diagonal elements of the
covariance matrix $K$ such that $\boldsymbol{\theta}$ becomes $(a,\lambda,\Gamma,P_{\text{GP}},s)$.

Using the Markov Chain Monte-Carlo (MCMC)
ensemble sampler \texttt{emcee} \citep{foremanmackey13} we sample the marginalized
posterior probability density functions (PDFs) of the five hyperparameters assuming 
uniform priors on the logarithm of each hyperparameter and maximizing the Gaussian
logarithmic likelihood function

\begin{equation}
\ln{\mathcal{L}} = -\frac{1}{2} \left( \mathbf{y}^T K^{-1} \mathbf{y}
+ \ln{\mathrm{det} K} + N \ln{2 \pi} \right),
\label{eq:like}
\end{equation}

\noindent where $\mathbf{y}$ is the vector of $N$ observations. In Model 1 $\mathbf{y}=$ 
the binned photometric data points\footnote{Binning
  the K2 photometry in one day bins results in $N=78$ compared to the 3439 unbinned photometric
  observations thus drastically increasing the computational efficiency of the evaluating Eq.~\ref{eq:like} in
  each step of the Markov chains.} whereas $\mathbf{y}=$ BIS time-series in Model 2.

\begin{figure}
\centering
\includegraphics[width=1.\linewidth]{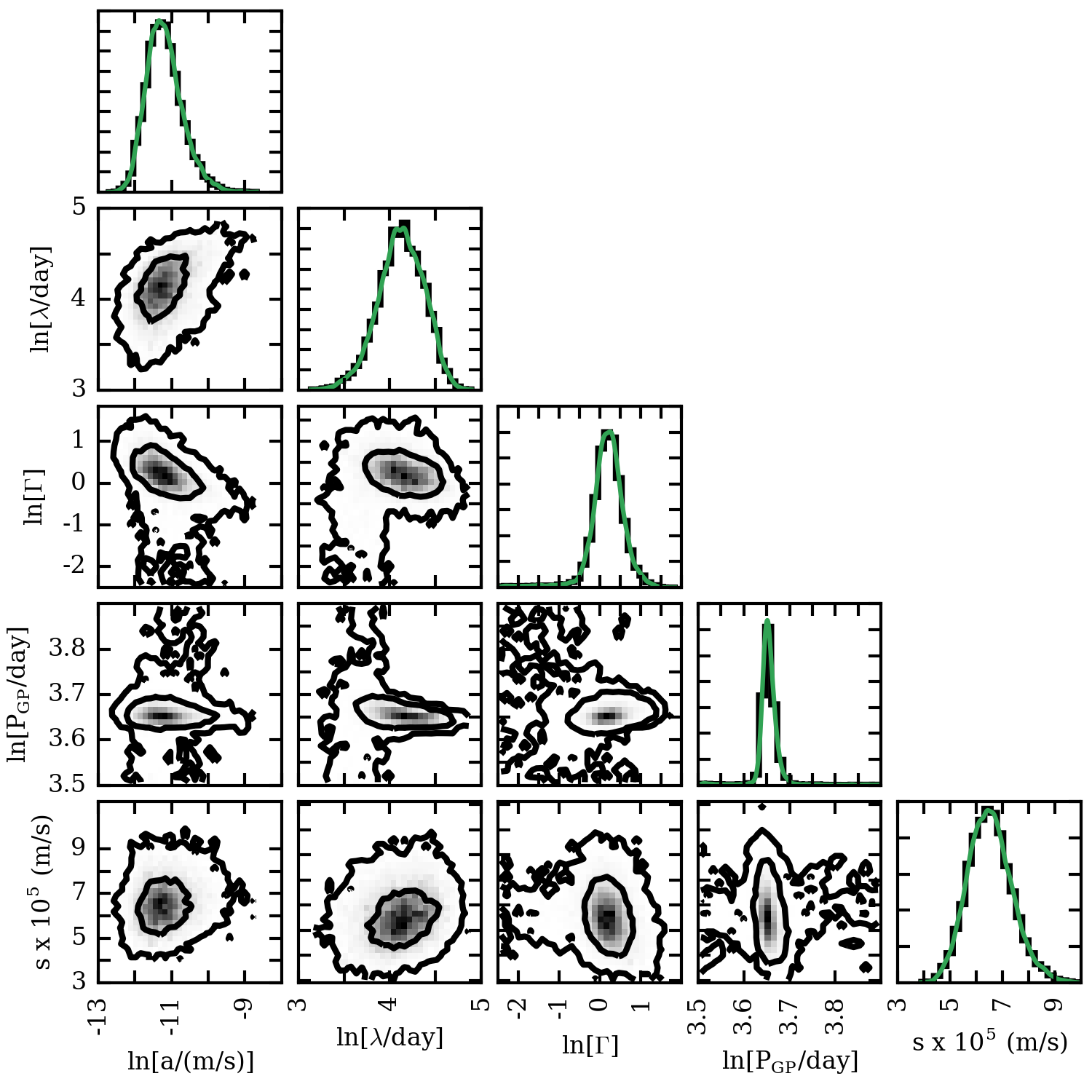}
\caption{The marginalized and joint posterior PDFs of the logarithmic GP hyperparameters
  used to model the K2 photometry shown in Fig.~\ref{fig:k2phot}.
  Kernel density estimations of each model parameter's posterior are overlaid on the
  histograms with \emph{solid green lines}. \label{fig:cornerGP}}
\end{figure}

The MCMC is initialized with 200 walkers and hyperparameter values
$(a,\lambda,\Gamma,P_{\text{GP}})= (max(\mathbf{y}-\langle \mathbf{y} \rangle), 10^2\text{ days}, 1, 36\text{ days})$.
We sample the logarithmic hyperparameters up to $\approx 10$ autocorrelation times to ensure
adequate convergence of the chains. We also monitor the acceptance fraction for each walker and insist
that it lies within $20-50$\%. The sampling of each hyperparameter's marginalized posterior PDF
commences following a burn-in phase of the same duration.
The resulting marginalized  and joint posterior PDFs are shown in Fig.~\ref{fig:cornerGP} along with
kernel density estimations of each marginalized 1D distribution. From the posterior
PDF of $P_{\text{GP}}$ we measure a stellar rotation period of \prot{} $= 38.6^{+0.6}_{-0.4}$ days.

\subsection{Joint modelling of RVs} \label{sect:jointmodel}
We proceed with modelling the RVs jointly with keplerian solutions for both K2-18b and c plus a trained
quasi-periodic GP to model the correlated RV residuals attributed to stellar jitter.
The marginalized posterior PDFs of the GP hyperparameters $\lambda,\Gamma$, and $P_{\text{GP}}$ from
training are
then used as informative priors in the joint RV analysis which treat the remaining GP hyperparameters
$a$ and $s$ as free parameters. We sample the informative priors using the kernel density estimations
of each hyperparameter's PDF obtained during training. This methodology
allows the model to learn the covariance structure of the stellar jitter through observations which are
independent of planetary sources and then apply that knowledge to the joint modelling of the RVs
thus distinguishing between stellar jitter and planet-induced Doppler shifts.

The RV modelling is again performed using \texttt{emcee}. In Models 1, 2, and 3
our RV model consists of 16 parameters including
the 5 GP hyperparameters discussed in Sect.~\ref{sect:gp}, the systemic velocity of K2-18 $\gamma_0$, the orbital periods
of the two planets $P$, their times of inferior conjunction $T_0$, their RV semi-amplitudes $K$,
and the MCMC jump parameters
$h=\sqrt{e}\cos{\omega}$ and $k=\sqrt{e}\sin{\omega}$ describing each planet's orbital eccentricity $e$
and argument of periastron $\omega$. This parameterization is chosen to minimize the correlation between
$e$ and $\omega$ as well as reduce the tendency for the MCMC sampler to favour high-eccentricity solutions
\citep{ford06}. In Model 4 we only consider 11 model parameters as no GP jitter model is included. 
Table~\ref{table:models} summarizes the adopted priors on each RV model parameter in each
of the models considered in this study. We adopt non-informative priors for all keplerian parameters other than the 
orbital period and time of mid-transit of K2-18b which are well-constrained by the transit light curve modelling in 
\citetalias{benneke17}.

\begin{table}
\tiny
\renewcommand{\arraystretch}{0.7}
\centering
\caption[]{Summary of RV Models and Adopted Priors}
\label{table:models}
\begin{tabular}{lc}
\hline \\ [-1ex]
Parameter & Prior \smallskip \\
\hline \\ [-1ex]
\multicolumn{2}{c}{Model 1} \\
\multicolumn{2}{c}{(2 planets + GP trained on $\mathbf{y}=$ K2 photometry)} \smallskip \\
\emph{GP hyperparameters} & \\
Covariance amplitude, $a$ [\mps{]} & $\mathcal{J}(0.1,30)^{\bullet}$ \\
Exponential timescale, $\lambda$ [days] & $p(\lambda|\mathbf{y})$ \\
Coherence, $\Gamma$ &  $p(\Gamma|\mathbf{y})$ \\
Periodic timescale, $P_{\text{GP}}$ [days] &  $p(P_{\text{GP}}|\mathbf{y})$ \\
Additive jitter, $s$ [\mps{]} & $\mathcal{J}(10^{-2},10)$ \smallskip \\ 
\emph{Keplerian parameters} & \\
$\gamma_0$ [\mps{]} & $\mathcal{U}(620,670)$ \\
$P_b$ [days] & $\mathcal{N}(32.939614,10^{-4})^{\circ}$ \\
$T_{0,b}$ [BJD-2,450,000] & $\mathcal{N}(7264.39144,6.3 \times 10^{-4})^{\circ}$ \\
$K_b$ [\mps{]} & $\text{\emph{mod}}\mathcal{J}(1,20)^{\ast}$ \\
$h_b = \sqrt{e_b}\cos{\omega_b}$ & $\mathcal{U}(-1,1)^{\dagger}$ \\
$k_b = \sqrt{e_b}\sin{\omega_b}$ & $\mathcal{U}(-1,1)^{\dagger}$ \\
$P_c$ [days] & $\mathcal{U}(8,10)$ \\
$T_{0,c}$ [BJD-2,450,000] & $\mathcal{U}(7259,7269)$ \\
$K_c$ [\mps{]} & $\text{\emph{mod}}\mathcal{J}(1,20)$ \\
$h_c = \sqrt{e_c}\cos{\omega_c}$ & $\mathcal{U}(-1,1)^{\dagger}$ \\
$k_c = \sqrt{e_c}\sin{\omega_c}$ & $\mathcal{U}(-1,1)^{\dagger}$ \medskip \\

\multicolumn{2}{c}{Model 2} \\
\multicolumn{2}{c}{(2 planets + GP trained on $\mathbf{y}=$ BIS)} \smallskip \\
see Model 1 & \\

\multicolumn{2}{c}{Model 3} \\
\multicolumn{2}{c}{(2 planets + untrained GP)} \smallskip \\
see Model 1  with the & \\
following modifications: & \\
ln Exponential timescale, $\ln{\lambda}$ [days] & $\mathcal{U}(-10,10)$ \\
ln Coherence, $\ln{\Gamma}$ &  $\mathcal{U}(-3,3)$ \\
ln Periodic timescale, $\ln{P_{\text{GP}}}$ [days] &  $\mathcal{U}(3.2,4)$ \medskip \\

\multicolumn{2}{c}{Model 4} \\
\multicolumn{2}{c}{(2 planets)} \smallskip \\
see keplerian parameters in Model 1 & \\

\hline
\end{tabular}
\begin{list}{}{}
\item {\bf{Notes.}} $^{(\bullet)} \mathcal{J}$ refers to a non-informative Jeffreys prior
  which is scale invariant; equal probability per decade which is necessary to sample
  multiple orders of magnitude \citep{gregory05}. \\
  $^{(\circ)}$ based on the transit light curve measurements from \citetalias{benneke17}. \\
  $^{(\ast)}$ $\text{\emph{mod}}\mathcal{J}(k,l)$ \mps{} 
  refers to a modified Jefferys prior on a parameter $A$ which behaves like a uniform
  prior for $A \ll$ the knee at k \mps{} and
  behaves like a Jeffreys prior at $A \gg k$ up to $l$. We use a modified Jeffreys prior on 
  the RV semi-amplitudes $K$ to sample multiple decades as a Jeffreys prior but also include
  $K=0$ \mps{} which a Jeffreys prior does not \citep{gregory05}. \\
  $^{(\dagger)}$ We further insist
  that $e = h^2 + k^2 < 1$.
\end{list}
\end{table}

\section{Results} \label{sect:results}
Here we compare results from the four considered RV models. Each model contains keplerian solutions for
each of the 2 planets. Additionally Model 1 models the RV residuals with a quasi-periodic GP regression model
that is trained on the K2 photometry in which the stellar rotation period is clearly detected. In this model
the stellar rotation period, and hence the GP periodic term, is sufficiently distinct from the orbital period
of K2-18b such that the two signals are not confused in our joint modelling and the measured semi-amplitude of
K2-18b is not mis-estimated. Model 2 models the RV residuals with a quasi-periodic GP regression
model that is trained on the BIS time-series which is contemporaneous with the RVs.
Model 3 models the RV residuals with an effectively unconstrained quasi-periodic GP and Model 4 neglects
any modelling of the RV residuals therefore assuming that they are uncorrelated.

\begin{figure*}
\centering
\includegraphics[width=1.\linewidth]{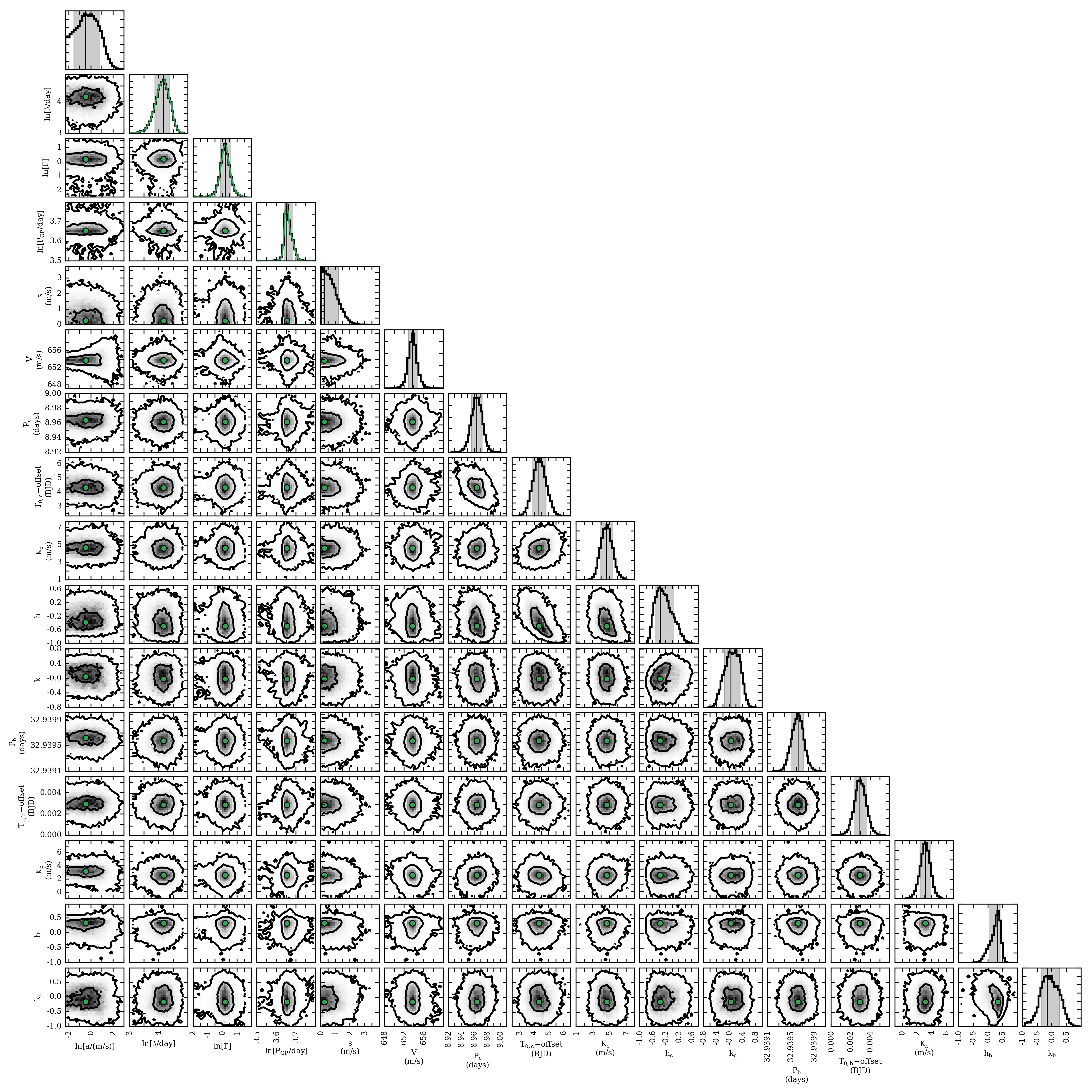}
\caption{The marginalized and joint posterior PDFs of the model parameters from Model 1 of
  the observed RVs. Model 1 of the observed RVs models the two planets with keplerian orbital
  solutions and the residual RV jitter with a GP regression model trained on the K2 photometry in
  Fig.~\ref{fig:k2phot}. Kernel density estimations of the trained posteriors are shown in the
  histograms of the logarithmic GP hyperparameters $\lambda, \Gamma$, and $P_{\text{GP}}$
  (columns 2,3,4). \label{fig:cornerRV}}
\end{figure*}

The \emph{maximum a-posteriori} (MAP) values of each model parameter along with the 16$^{\text{th}}$
and $84^{\text{th}}$ percentiles of
the marginalized posterior PDFs are reported in Table~\ref{table:k2184}. The marginalized and joint posterior
PDFs of the model parameters used in Model 1 are shown in Fig.~\ref{fig:cornerRV}. In Sect.~\ref{sect:modelcomp}
we shall see that Model 1 is the best predictor of the observed RVs and therefore we only show the results from
Model 1 in Fig.~\ref{fig:cornerRV}.

We emphasize that \emph{all} keplerian 
model parameters for the two planets are consistent at $1\sigma$ across all four models. Recall that the stellar
rotation period is only well-constrained in Model 1 via the K2 photometry and yet the measured semi-amplitudes of
K2-18b are consistent in each of the four models. This further demonstrates that there is minimal confusion between
the RV signals at the stellar rotation period (38.6 days) and at the orbital period of K2-18b ($\sim 32.93$ days)
both of which are not distinctly detected in the periodogram of the raw RVs (see top panel of Fig.~\ref{fig:periodograms})
but appear to be hidden within a forest of peaks spanning periodicities between $\sim 25-45$ days. 
The consistency of all measured keplerian parameters in each model also suggests that the RV residuals,
following the removal of the two MAP keplerian solutions, are weakly correlated because nearly identical RV solutions
are obtained with and without a GP treatment of the RV residuals following the removal of our planet models.
That is that K2-18 appears to be a spectroscopically quiet star with the majority of its observed RV variation being
attributable to planetary companions. Being spectroscopically quiet is promising for the prospect of
transmission spectroscopy of K2-18b; an observation that is significantly complicated by the presence of
stellar jitter. The quiet nature of K2-18 is highlighted by its low measured value of $\log{R'_{\text{HK}}}=-5.247$.  

\subsection{RV Model Comparison} \label{sect:modelcomp}
A formal model comparison between each of the four considered models is performed using
\emph{time-series cross-validation}
to compute the likelihood of each model given various training and testing subsets of the 
observed RVs \citep{arlot10}.
We split the RVs into chronological training sets with sizes ranging from 20 measurements to the
size of the full dataset less one (i.e. 74 measurements). The model parameters for each of the
four considered models are optimized on the training
set and the likelihood of the corresponding model is evaluated on the testing set. The testing set is simply
the next observation chronologically following the final observation in the training set. The resulting median likelihood
and median absolute deviation for each model is
reported at the bottom of Table~\ref{table:k2184} and is used to distinguish which of our four
RV models performs optimally on the prediction of unseen RV measurements and thus best fits the data without over-fitting.
Through time-series cross-validation we find that Model 1 is the best predictor of the observed RVs. In the remainder of
this study we will consider the results from Model 1 to be the measured values of the planets K2-18b and c.

To confirm that we have detected a second planet K2-18c in our RV data, we perform a second round of time-series
cross-validation calculations. In these calculations we will compare
three RV models each containing 0, 1, or 2 planets. We consider K2-18b to be the
lone planet in the one planet model. In each model we also consider
a GP jitter model trained on the K2 photometry similarly to Model 1 above.
Following the same methodology as previously
discussed we find median logarithmic likelihoods of $\ln{\mathcal{L}_0}=-2.693 \pm 0.056$,
$\ln{\mathcal{L}_1}=-2.642 \pm 0.047$, and $\ln{\mathcal{L}_2}=-2.566 \pm 0.026$.
From this we find that $\ln{\mathcal{L}_2}-\ln{\mathcal{L}_1} = 0.076 \pm 0.054 > 0$ therefore arguing that
the two planet model is the best predictor of unseen RV measurements and confirming that our two planet model
containing both K2-18b and c is the RV model most favoured by the data. 

\begin{figure}
  \centering
  \includegraphics[width=\hsize]{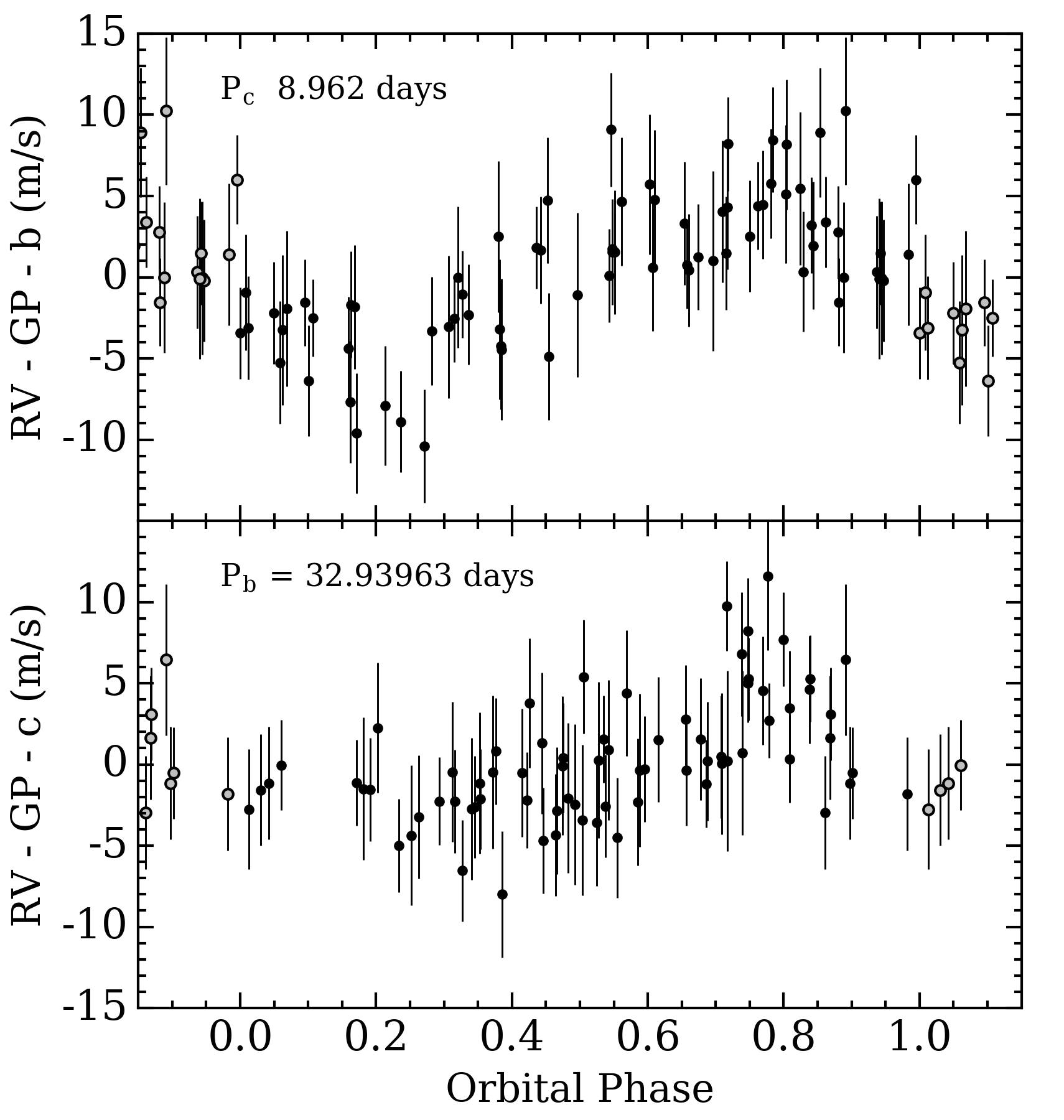}%
   \hspace{-\hsize}%
  \begin{ocg}{fig:curveoff}{fig:curveoff}{0}%
  \end{ocg}%
  \begin{ocg}{fig:curveon}{fig:curveon}{1}%
   \includegraphics[width=\hsize]{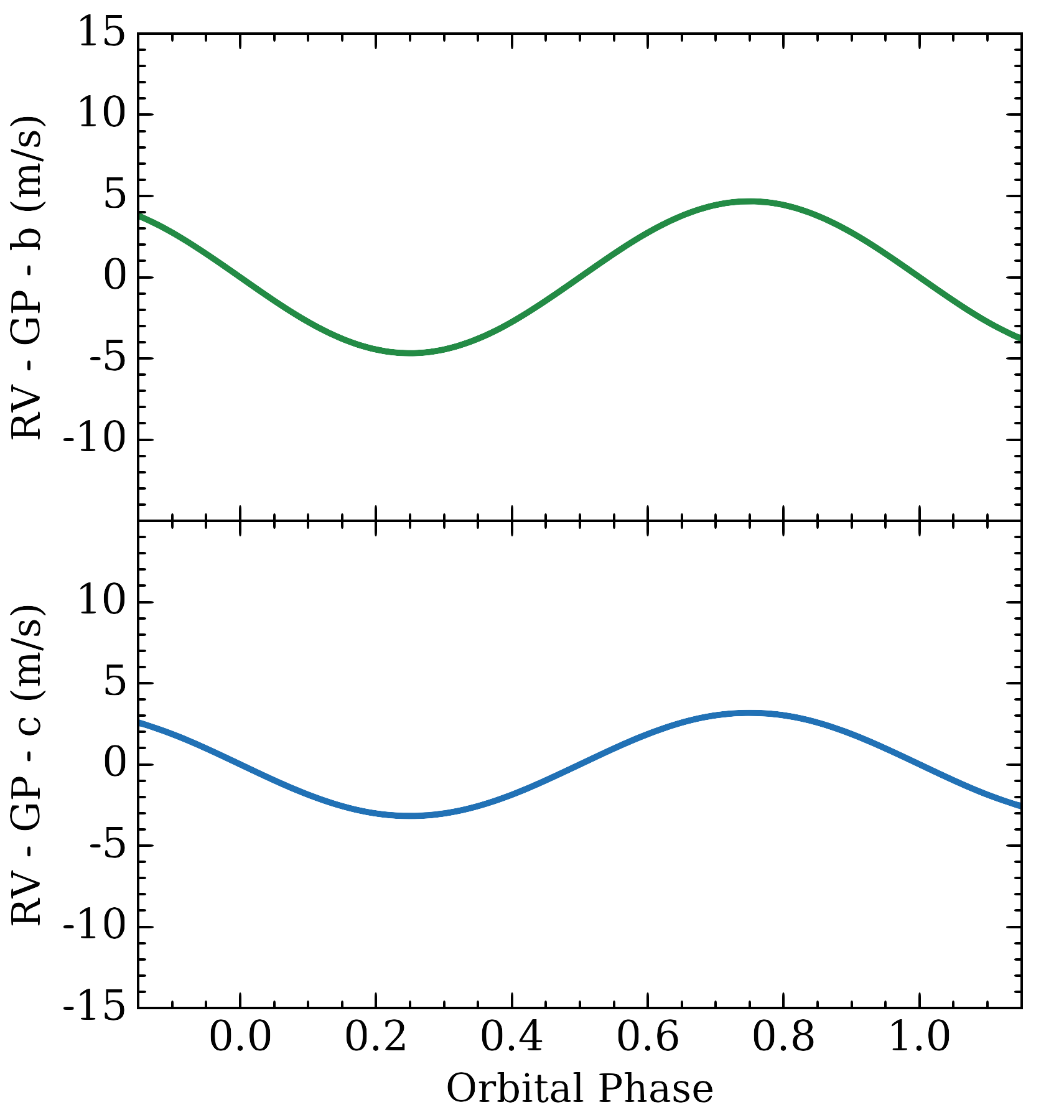}%
  \end{ocg}
  \hspace{-\hsize}%
  \caption{The phase-folded RVs for each planet in the K2-18
  planetary system (\emph{top}: K2-18c and \emph{bottom}: K2-18b). The RVs have
  been corrected for stellar jitter with a quasi-periodic GP
  model trained on the K2 photometry.
  The \ToggleLayer{fig:curveon,fig:curveoff}{\protect\cdbox{\emph{solid curves}}}
  depict the maximum a-posteriori keplerian orbital solutions for each planet
  with fixed circular orbits.}
  \label{fig:rvs}
\end{figure}

The Model 1 MAP keplerian orbital solutions for K2-18b and c, but with eccentricities fixed to zero,
are shown in Fig.~\ref{fig:rvs}. The RV data are phase-folded to each planet's MAP orbital period and
time of inferior conjunction and are corrected for stellar jitter based on the mean GP jitter model trained
on the K2 photometry. The residual rms following the removal of all modelled contributions is 2.89
  \mps{.} This is less than the median photon noise limit of the measured RVs of 3.56 \mps{} suggesting that
  we have modelled all significant RV contributions. For comparison, the residual rms achieved in Model 4, which
  neglects any red noise modelling following the removal of the two keplerian solutions, is 3.16 \mps{} which is
  also less than the median photon noise limit suggesting that the GP regression modelling alone in Models 1,2 and 3
  do not result in over-fitting of the data.

\section{Searching for transits of K2-18c} \label{sect:transit}
From our RV analysis in Sect.~\ref{sect:results} we derived the approximate linear ephemeris of
K2-18c. We can therefore predict the passage of K2-18c at inferior conjunctions within the
mostly continuous K2 photometric monitoring shown in Fig.~\ref{fig:k2phot}. The green ticks in
Fig.~\ref{fig:k2phot} depict the 9 such passages of K2-18c. Given the comparable minimum masses of
K2-18b and c, it is reasonable to expect that the two planets also have comparable radii (recall
$r_{p,b} \sim 2.28$ R$_{\oplus}$). Furthermore \citep{ciardi13} argued that Kepler multi-planet
systems with planet radii $\lesssim 3$ R$_{\oplus}$ do not exhibit a size---semi-major axis
correlation such that the inner K2-18c is not expected to have undergone significant atmospheric
escape compared to K2-18b. However the two $10\sigma$ transits of K2-18b
are clearly discernible by-eye in the K2 photometry whereas the predicted transits of K2-18c
are not. This suggests that either K2-18c is much smaller than K2-18b such that its resulting
transit depth is below the threshold for detection or that
the orbit K2-18c is mutually inclined with that of K2-18b such that it misses a transit configuration.

Here we attempt to confirm that K2-18c is indeed not transiting in the K2 light curves. To do so
we perform an MCMC sampling of the K2-18c transit
model parameters ($P_c$, $T_{0,c}$, $r_{p,c}/R_s$, $a_c/R_s$, and impact parameter $b_c$)
using the K2 photometry and
following the removal of the known transits of K2-18b. A quadratic limb-darkening law is
assumed with fixed parameters in the Kepler bandpass: $a=0.3695$ and $b=0.3570$. These values
are interpolated from the tables of
\cite{claret11} based on the known K2-18 surface gravity and effective temperature. 
In each MCMC step we compute the corresponding transit model using the \texttt{batman} implementation
\citep{kreidberg15} of the \cite{mandel02} transit model.
We assume a circular orbit of K2-18c and adopt the same MCMC methodology
utilized on the RV data in Sect.~\ref{sect:jointmodel}. The orbital period and time of inferior
conjunction (i.e. time of mid-transit) are sampled from their joint RV posterior which maintains
their apparent correlation (see Fig.~\ref{fig:cornerRV}). Priors on
the scaled planet radius, and impact parameter are assumed uniform. In this way
the scaled planetary radius is uncorrelated with its measured minimum mass and the impact parameter is
constrained to be $|b_c|<1$ as is required for a transit to occur.

Based on our MCMC analysis we find that the values of $r_{p,c}/R_s$ are consistent with zero
i.e. no transit is detected in the K2 data.
Assuming the most-likely value of $R_s = 0.411$ R$_{\odot}$ 
we calculate a planet radius upper limit of $r_{p,c} < 0.52$ R$_{\oplus}$ at 99\% confidence.
If K2-18c were this size and  
transiting, albeit unknowingly due to its small radius, the planet would have a bulk density
of $\gtrsim 295$ g cm$^{-3}$ or $\gtrsim 54$ $\rho_{\oplus}$; an unphysically large value given
the compressibility of pure iron. Thus
we conclude that K2-18c is not transiting in the K2 data and is therefore not co-planar with
K2-18b despite having a smaller orbital separation.

\begin{figure}
  \centering
  \includegraphics[width=\hsize]{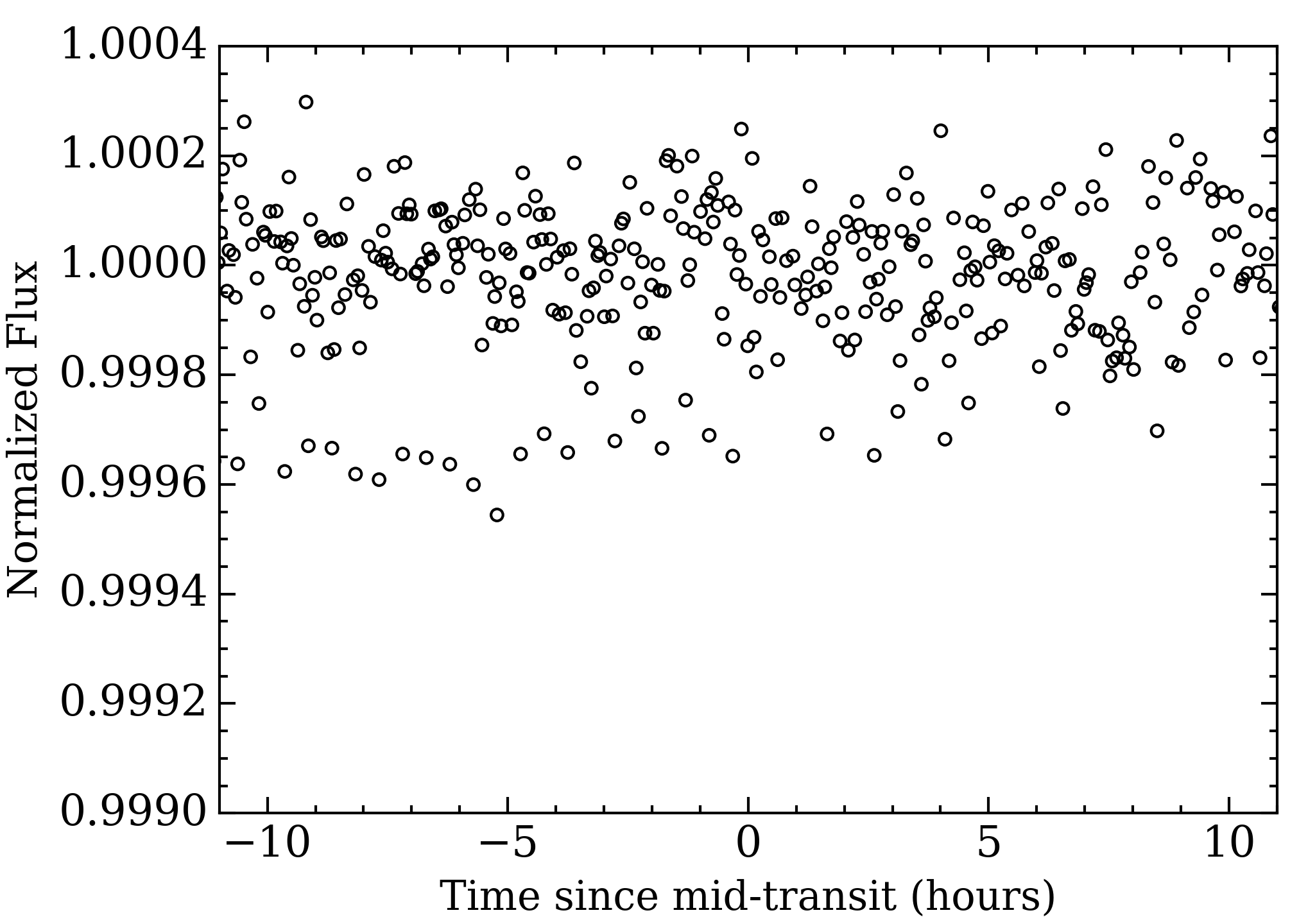}%
   \hspace{-\hsize}%
  \begin{ocg}{fig:1off}{fig:1off}{0}%
  \end{ocg}%
  \begin{ocg}{fig:1on}{fig:1on}{1}%
    \includegraphics[width=\hsize]{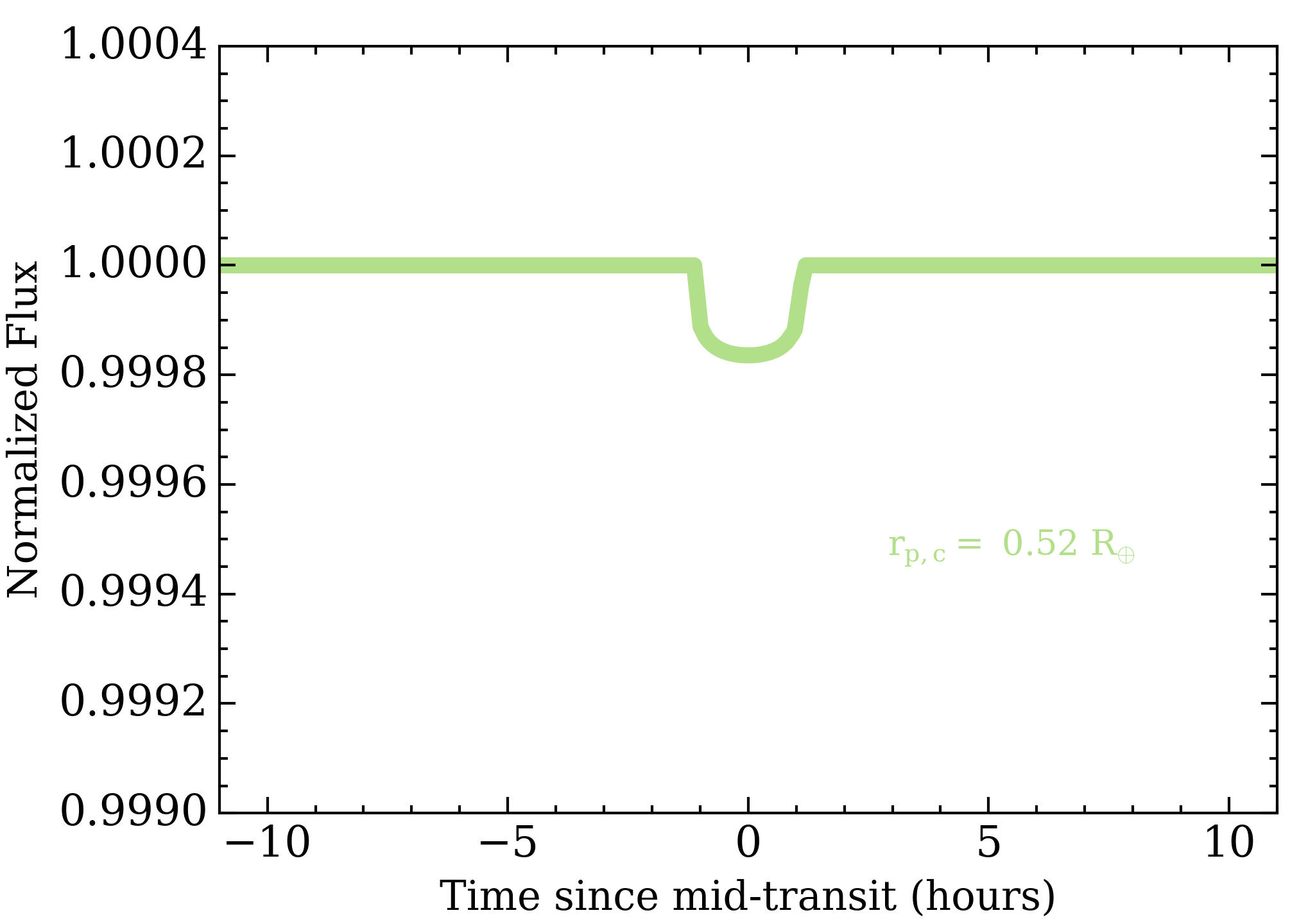}%
  \end{ocg}
  \hspace{-\hsize}%
  \begin{ocg}{fig:2off}{fig:2off}{0}%
  \end{ocg}%
  \begin{ocg}{fig:2on}{fig:2on}{1}%
    \includegraphics[width=\hsize]{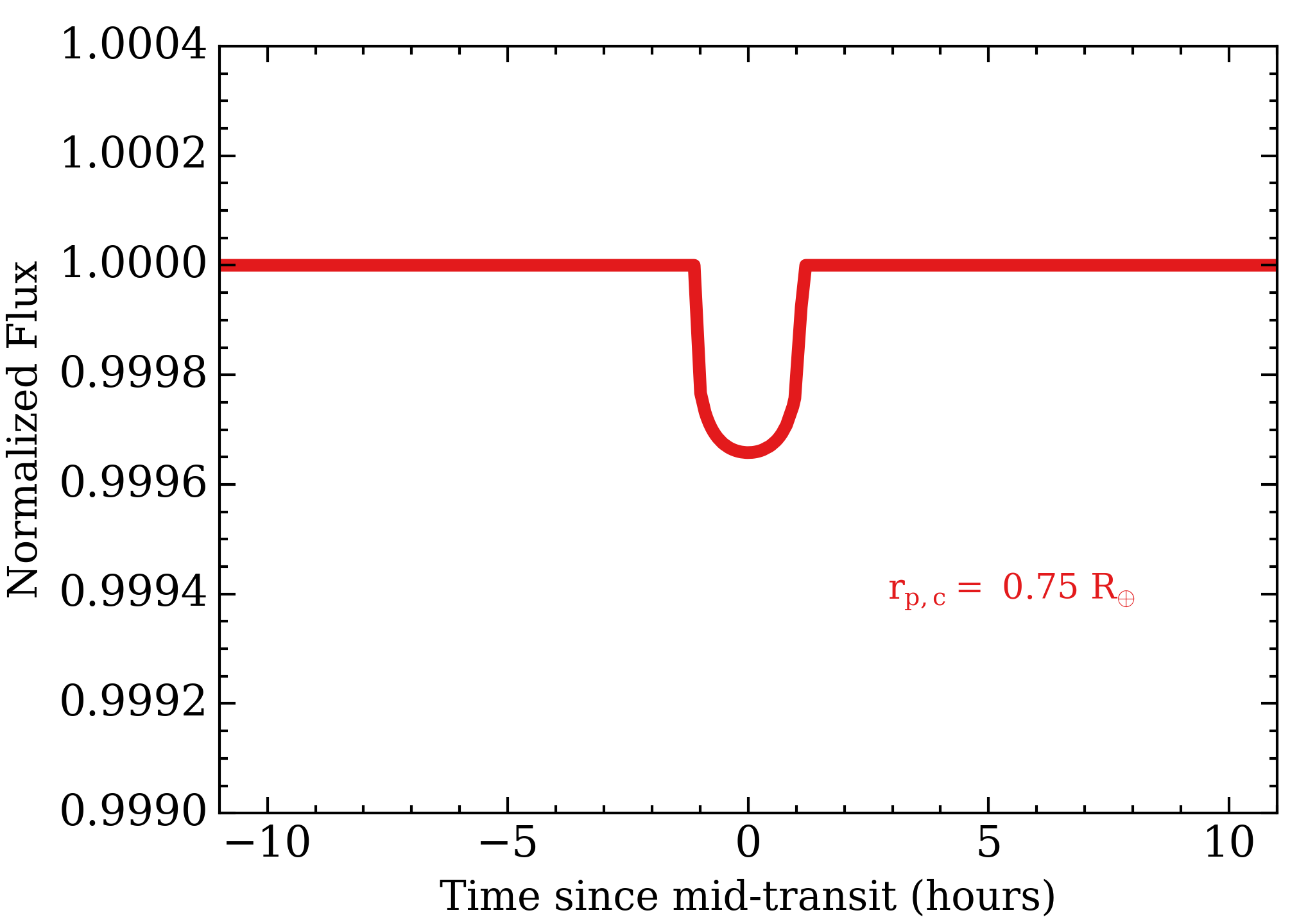}%
  \end{ocg}%
  \hspace{-\hsize}%
  \begin{ocg}{fig:3off}{fig:3off}{0}%
  \end{ocg}%
  \begin{ocg}{fig:3on}{fig:3on}{1}%
    \includegraphics[width=\hsize]{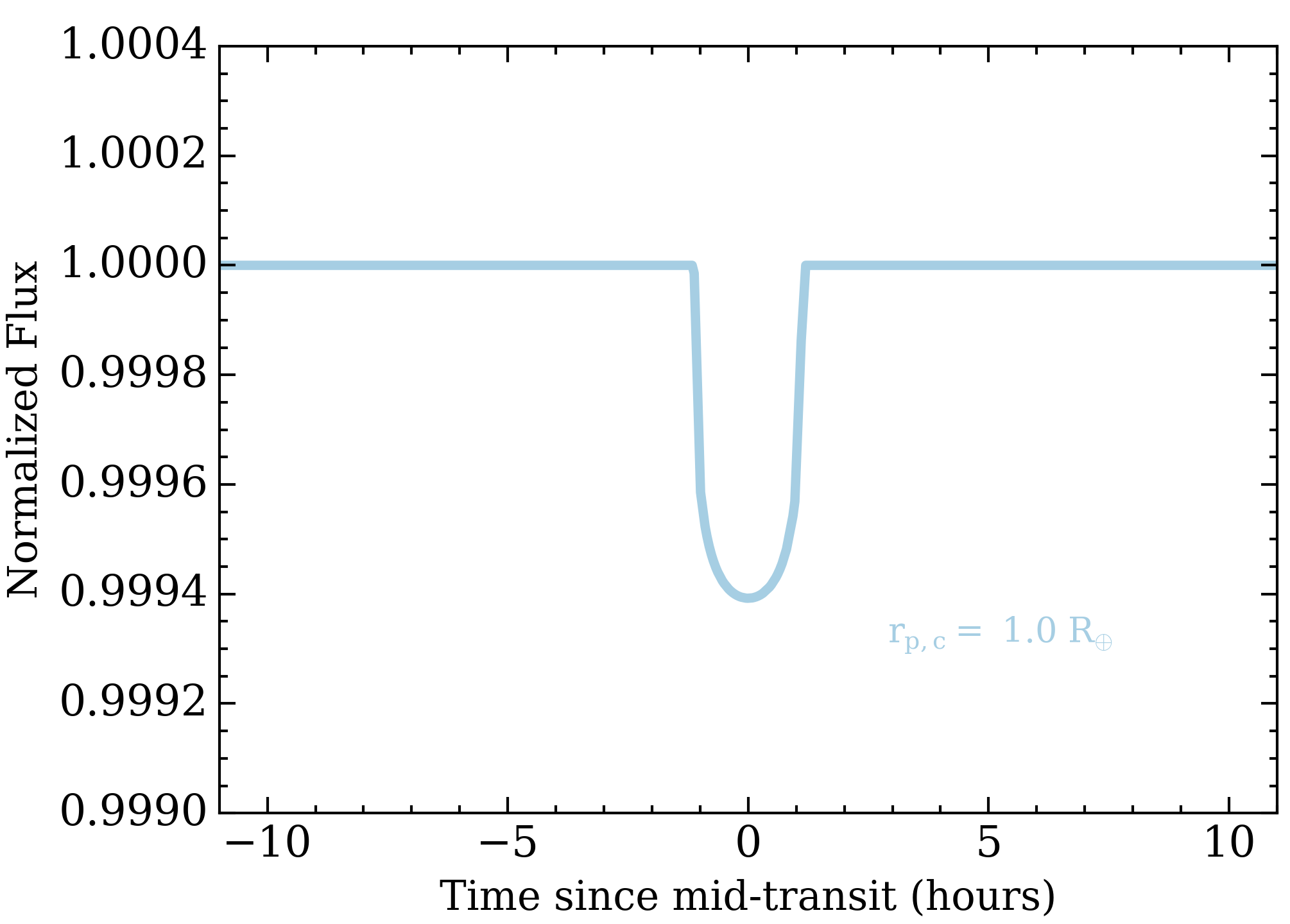}%
  \end{ocg}%
  \hspace{-\hsize}%
  \begin{ocg}{fig:4off}{fig:4off}{0}%
  \end{ocg}%
  \begin{ocg}{fig:4on}{fig:4on}{1}%
    \includegraphics[width=\hsize]{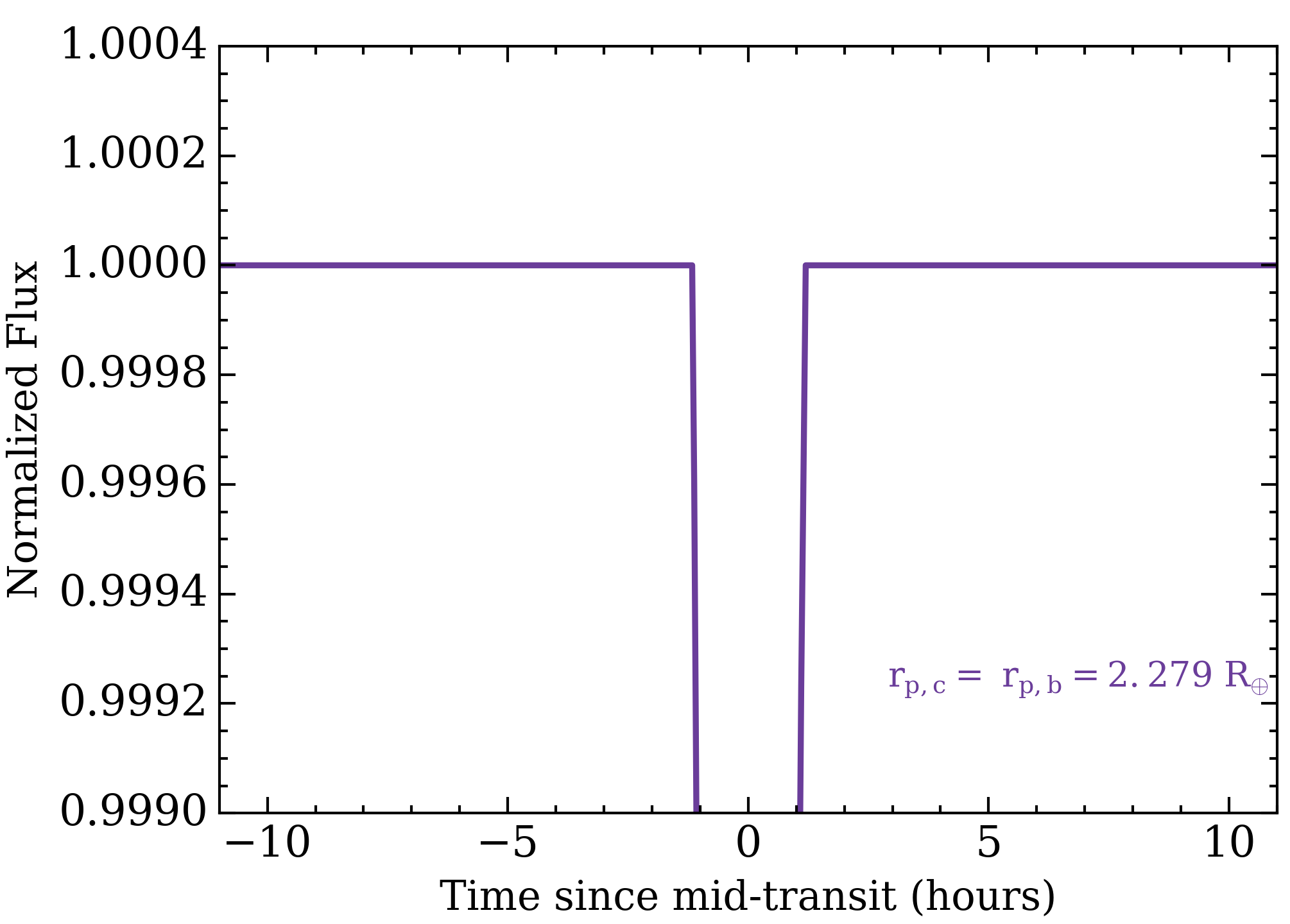}%
  \end{ocg}%
  \caption{The K2 light curve of K2-18 phase-folded to the maximum a-posteriori orbital period
    and time of mid-transit of K2-18c from Model 1. Four transit light curve models are over-plotted
    for various illustrative values of the assumed size of K2-18c: the 99\% upper limit on $r_{p,c}$   
    \ToggleLayer{fig:1on,fig:1off}{\protect\cdbox{(0.52 R$_{\oplus}$)}}, 
    \ToggleLayer{fig:2on,fig:2off}{\protect\cdbox{0.75 R$_{\oplus}$}},
    \ToggleLayer{fig:3on,fig:3off}{\protect\cdbox{1 R$_{\oplus}$}}, and the size of K2-18b
    \ToggleLayer{fig:4on,fig:4off}{\protect\cdbox{(2.279 R$_{\oplus}$)}}. No transit of K2-18c is
    detected in the data.}
  \label{fig:transit}
\end{figure}

To visualize the data a selection of light
curve models are compared to the phase-folded K2 photometry in Fig.~\ref{fig:transit}. Models shown
include a suite of $r_{p,c}$ values including its upper limit derived from MCMC (0.52 R$_{\oplus}$),
0.75 R$_{\oplus}$, 1 R$_{\oplus}$, and the radius of K2-18b (2.279 R$_{\oplus}$) which is detected at
the 10$\sigma$ level in the K2 photometry \citep{montet15}. 

\section{Dynamical Stability and Eccentricity Restrictions} \label{sect:dynam}
The non-detection of K2-18c in transit (see Sect.~\ref{sect:transit}) suggests that its orbital plane
is non-coplanar with the outer
transiting K2-18b whose semi-major axis is $\sim 2.4$ times greater than K2-18c's. The orbital inclination
of K2-18b is $89.5785^{+0.0079}_{-0.0088}$ degrees with a corresponding impact parameter
of $0.601^{+0.013}_{-0.011}$ \citepalias{benneke17}. In order for the orbit of K2-18c
to not pass in front of its host star its orbital inclination must be tilted
either $\gtrsim 1.41^{\circ}$ or $\lesssim -2.25^{\circ}$ from the orbit of K2-18b depending on
which hemisphere of the stellar disk its transit chord will traverse. 
Such a mutual inclination is consistent with the peak in the distribution of Kepler multi-planet
mutual inclinations \citep{figueira12, fabrycky14}.
If indeed the planetary angular momentum vectors are within only a few degrees and therefore
nearly aligned then we can analytically evaluate their Hill stability
given estimates of their orbital eccentricities and assuming an inclination
correction factor that is close to unity \citep{gladman93}. If we assume the simplest case of 
initially circular orbits then the system is strongly Hill stable given that the two planets
are currently separated by $\sim 23$ mutual Hill radii. 


Accurate orbital eccentricities of small planets with precision radial velocities are notoriously
difficult to measure. For example the change in RV semi-amplitude of a circular K2-18b compared to an
eccentricity of 0.1 is $\lesssim 2$ cm s$^{-1}$ ($\sim 0.5$\% of $K_b$) or 15 cm s$^{-1}$
($\sim 5$\% of $K_b$) for an eccentricity of 0.3. The aforementioned values are both at least an
order of magnitude smaller than the characteristic RV uncertainty of the HARPS measurements
presented in this work. Given that the system is Hill stable at small eccentricities 
we can use dynamical simulations to constrain the orbital eccentricities of the
planets insisting that the system remain stable throughout its simulated evolution. 

To constrain the planet eccentricities
we perform a suite of $10^4$ dynamical integrations wherein we sample linearly
each planet's $e \in [0,1)$.
In each simulation the orbital inclination of K2-18b is drawn from
$\mathcal{N}(89.5785^{\circ},0.0084^{\circ})$ while the
system's mutual inclination is drawn from $\mathcal{N}(\Delta i_{\text{min},c},1.5^{\circ})$ 
such that the planet inclinations remain uncorrelated with the orbital eccentricities thus permitting
an unbiased assessment of the system's stability across the keplerian parameter space. 
We insist that K2-18c be non-transiting at the start of each simulation by setting
$\Delta i_{\text{min},c}$ to be the minimum mutual inclination required for $|b_c|>1$
and rejecting draws for which this is not true. 
The dispersion in sampled mutual inclinations is tuned such that the mode of the resulting distribution
lies within $\sim 1-2^{\circ}$ \citep{fabrycky14}.
Each planet's initial semi-major axis, true anomaly, and absolute mass is drawn from a Gaussian
distribution with a mean value equal to the parameter's MAP value from Model 1 in
Table~\ref{table:k2184} and with standard deviations equal to its average measurement uncertainty.
The stellar mass is drawn from $\mathcal{N}(0.359,0.047)$ M$_{\odot}$.
The ascending node longitudes and arguments of
periapsis are both drawn from $\mathcal{U}(-\pi,\pi)$. The system is then integrated forward in time
from the epoch of the first K2 photometric observation (BJD=2456810.26222) for $10^6$ years using 
the Wisdom-Holman symplectic integrator \texttt{WHFast} \citep{rein15} implemented in the
open-source \texttt{REBOUND} N-body package \citep{rein12}. These integrations are not intended to
provide a comprehensive overview of the system's dynamical stability but rather are 
useful to show that the 
system can remain stable up to at least 1 Myr and provide constraints on the planet
eccentricities.

\begin{figure}
\centering
\includegraphics[width=.95\linewidth]{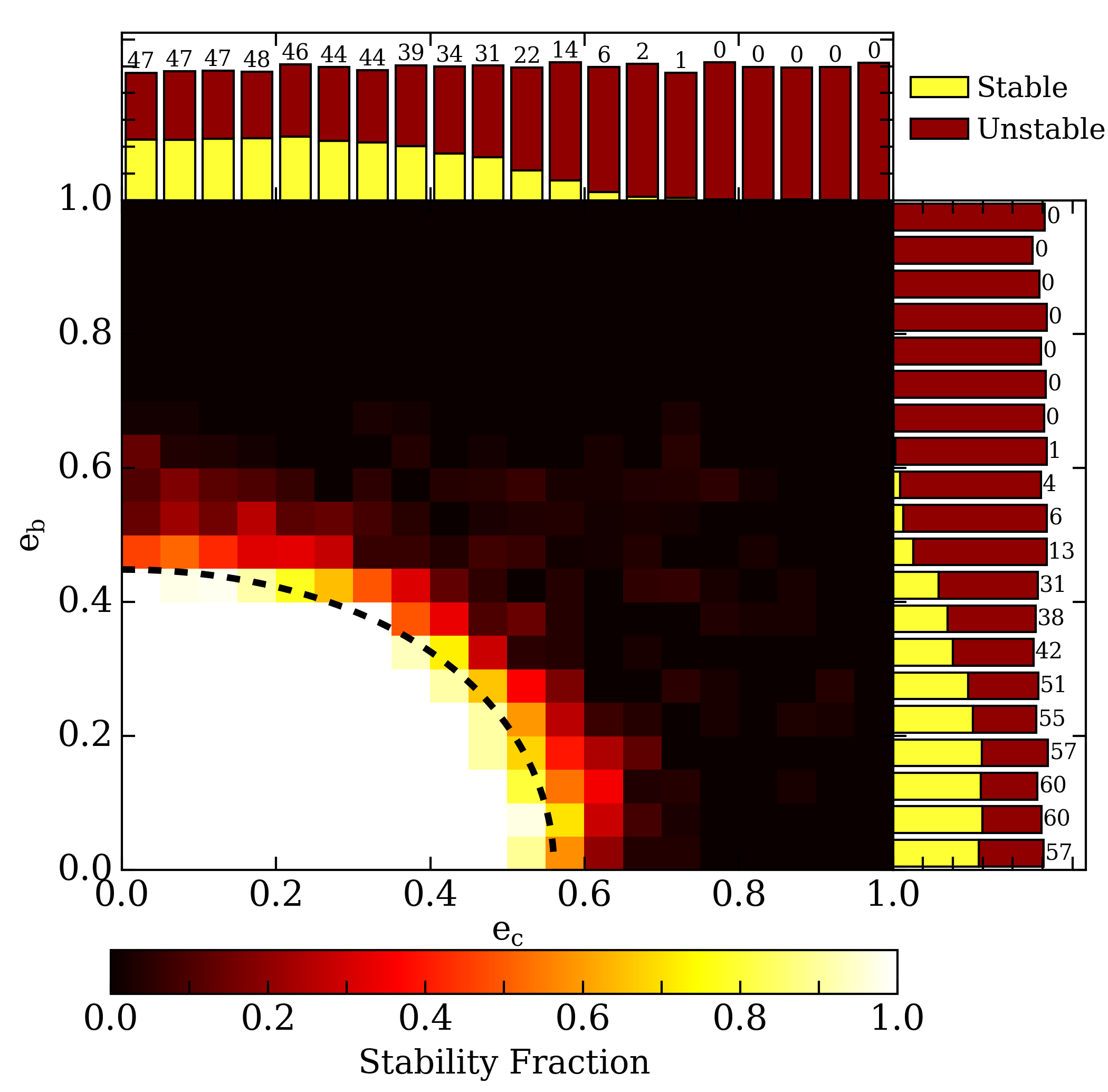}
\caption{Smoothed 2D map depicting the fraction of stable systems as a function of each planet's
  eccentricity based on a suite of dynamical integrations.
  The \emph{dashed black curve} depicts the analytic condition for Hill stability from
  \cite{gladman93} assuming the MAP masses and semi-major axes from Model 1 in
  Table~\ref{table:k2184}. 1D histograms depict the number of
  stable and unstable (\emph{yellow} and \emph{red} respectively) systems in
  eccentricity bins of width 0.05 and marginalized over all other dynamical parameters.
  The annotated numbers report each bin's stability fraction in percentages. 
  \label{fig:dynam}}
\end{figure}

We classify stable integrations as those in which the minimum distance between the planets never
becomes less than their mutual Hill radius. The fraction of stable systems as a function of each
planet's eccentricity and marginalized over all other dynamical parameters 
is shown in Fig.~\ref{fig:dynam}. Strong correlations between the fraction of stable systems and
dynamical parameters other than planet eccentricities was not apparent so we focus here on the
effect of eccentricities only. At small eccentricities there is a 
large stable region wherein the fraction of systems that remain stable is $\gtrsim 80$\%
and the system is known to be Hill stable based on the analytic criterion.
As we increase either planet's eccentricity the fraction of stable systems decreases.
This is also illustrated by further marginalizing over planet eccentricities and considering the 
1-dimensional representations of each system's stability fraction in the histograms shown in 
Fig.~\ref{fig:dynam}.

The RV analysis discussed Sect.~\ref{sect:results} and our dynamical simulations provide two
independent methods for constraining the orbital eccentricities of the K2-18 planets. We can
therefore combine these independent results by using the dynamical stability fractions
shown in Fig.~\ref{fig:dynam} as an additional prior on the $i^{\text{th}}$ planet's derived
eccentricity posterior: $e_i = h_i^2 + k_i^2$. To do this we resample each planet's RV eccentricity
posterior and accept draws with a probability equal to the stability fraction at that drawn
eccentricity value $\pm$ 0.025. This choice of bin width was varied between 0.01 and 0.1 and
was found not to have a significant effect on the results. In this way numerous random samples
from each planet's RV eccentricity posterior
are rejected due to the low corresponding stability fraction. This is especially true
for large eccentricities wherein the system no longer satisfies the Hill stability criterion.
From the modified eccentricity posteriors we can calculate the
$99^{\text{th}}$ percentiles and find that $e_b < 0.43$ and $e_c < 0.47$   
at that confidence level. These are the eccentricity values reported in Table~\ref{table:k2184}
and represent a more stringent evaluation of each planet's eccentricity than considering the RV data
alone.

\section{Discussion} \label{sect:disc}
With a set of 75 precision radial velocity measurements taken with the HARPS spectrograph
we have obtained a robust mass measurement of the transiting HZ super-Earth K2-18b and
detected a second super-Earth K2-18c. The orbit of the newly discovered K2-18c lies
interior to that of K2-18b and yet the planet is non-transiting. This implies that the orbital
planes of the planets are mutually inclined. In order for K2-18c
to not be seen in-transit the planetary system requires a mutual inclination of just
$\gtrsim 1.4^{\circ}$ which is consistent with the observed distribution of mutually inclined
multi-planet systems \citep{figueira12, fabrycky14}. Dynamical simulations of the system
revealed that the oscillation timescale of the planets' orbital inclinations is
$\mathcal{O}(10^6 \text{ years})$ suggesting that it may take many years before K2-18c reaches a transiting
configuration. Although exactly how long depends sensitivity on its current inclination which remains
unknown. The discovery of RV planets in transiting M dwarf planetary systems 
further emphasizes the prevalence of multiple Earth to super-Earth-sized planets around
nearby M dwarfs and that these additional planets can be uncovered with moderate RV follow-up
\citep{cloutier17}. Multi-planet systems such as K2-18 provide unique opportunities to study
planet formation processes around M dwarfs via direct comparative planetology. 

The presence of a second planet in the K2-18 transiting system will result in mutual 
planetary interactions thus making the orbit of K2-18b non-keplerian and possibly resulting
in an observable transit timing variation (TTV). Assuming a mutual inclination of K2-18c
that just misses a transiting configuration, we estimate the expected TTVs 
of K2-18b using the \texttt{TTVFaster} package \citep{deck14, agol16}. We adopt the maximum a-posteriori
masses and orbital periods of the two planets from Model 1 and uniformly sample their
eccentricities up the 99\%
upper limits reported in Table~\ref{table:k2184}. The remaining orbital parameters of the planets that
are unconstrained by the RV data 
are sampled uniformly between 0 and $2\pi$. We find a maximum TTV for K2-18b of $\sim 40$
seconds which is slightly less than, but of the same order as the uncertainty in its measured
time of mid-transit \citepalias[$\sim 50$ seconds;][]{benneke17}.
Thus with photometric monitoring of at least comparable quality to the K2 photometry
shown in Fig.~\ref{fig:k2phot}, detecting TTVs in the K2-18 multi-planet system is unlikely to provide
any significant new insight into the nature of the system. Indeed no significant TTVs were observed in
the Spitzer light curves from \citetalias{benneke17}.

\begin{figure}
\centering
\includegraphics[width=1.\linewidth]{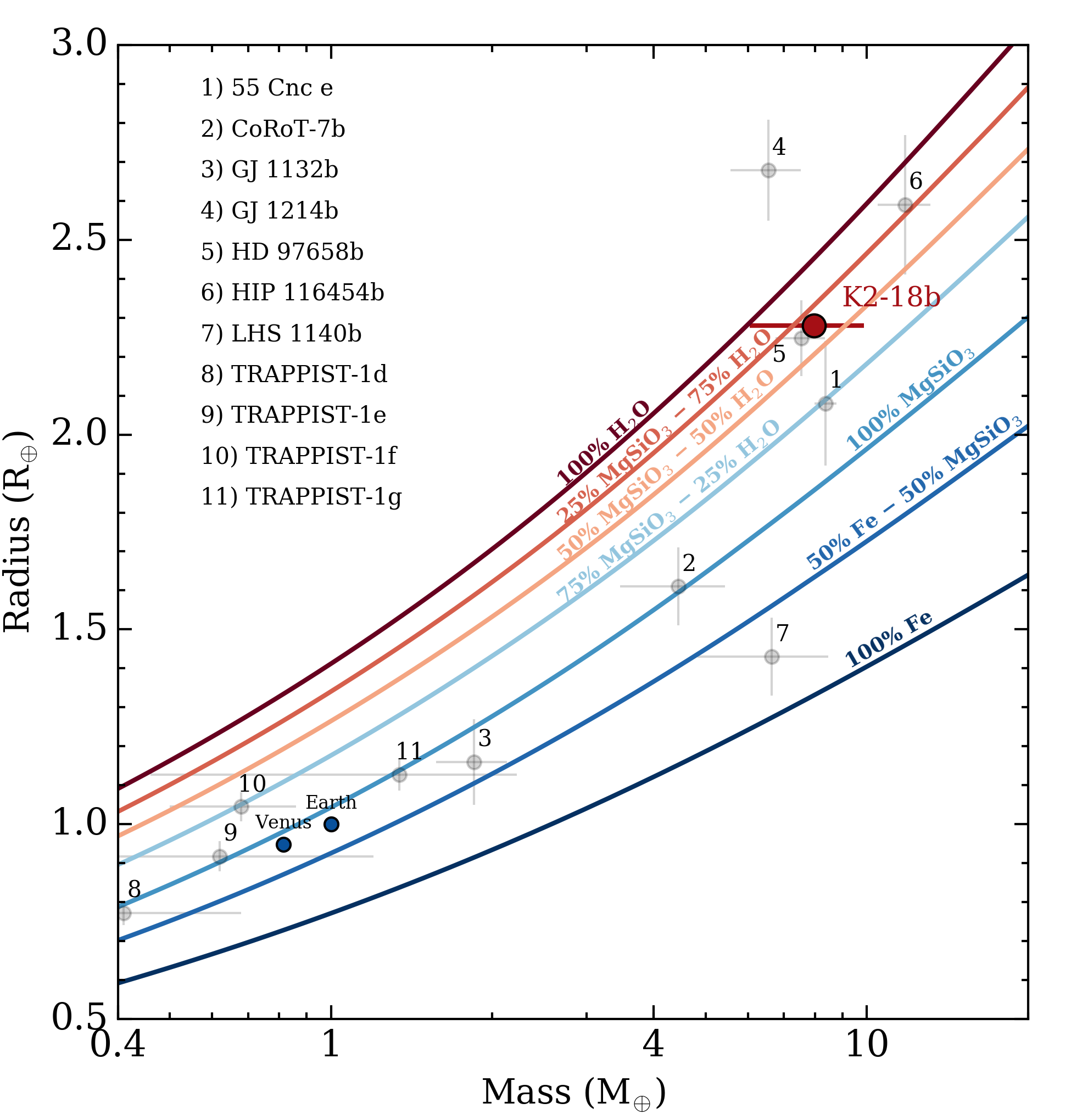}
\caption{K2-18b along with a sample of other small exoplanets in the planetary mass
  and radius space. Overlaid curves are two-component interior structure models of
  fully-differentiated solid planets with mass fractions annotated for each curve.
  \label{fig:mr}}
\end{figure}

With our measured mass of K2-18b the planet joins a select group of HZ planets with constraints
on both its mass and radius. This represents a significant step towards searching for potentially habitable
planets around stars earlier than $\sim$ M4 \citep{dittmann17}. 
With its maximum a-posteriori mass of $7.96 \pm 1.91$ M$_{\oplus}$
the bulk density of K2-18b ($\rho_{p,b} = 3.7 \pm 0.9$ g cm$^{-3}$)
lies between that of an Earth-like rocky planet and a
low density Neptune-like planet. The planet is therefore likely too large to be a terrestrial
Earth-like planet \citep{valencia07, fulton17}. Including K2-18b on the exoplanet mass-radius diagram
in Fig.~\ref{fig:mr} we find that the internal structure of K2-18b is
consistent with a range of two-component solid-planet models \citep{zeng13}
owing to the uncertainty in its measured mass which is at the level of $\sim 24$\%.
In particular the $1\sigma$ lower mass limit of K2-18b permits it to be a
complete `water-world' whereas its upper mass limit is consistent with a rock-dominated
interior surrounded by a significant mass fraction of water ice. 
In this parameter space the physical parameters of K2-18b are most similar to the super-Earth
HD 97658b \citep{vangrootel14} despite receiving $\sim 65$ times less insolation.
Furthermore K2-18b is of a similar mass to the habitable zone planet LHS 1140b \citep{dittmann17}
and receives a comparable level of insolation despite being $\sim 1.6$ times larger than LHS
1140b. Analyzing the mass-radius relationship of these small planets over a range of equilibrium
temperatures is a critical step towards
understanding which of these systems have retained significant atmospheric content
thus making them more suitable to extraterrestrial life.

Distinguishing between K2-18b as a pure water-world or a scaled-up Earth with a significant gaseous
envelope will likely require transmission spectroscopy follow-up observations either with high-resolution
spectrographs from the ground 
or from space with JWST. With $J$=9.8, $H$=9.1, and $K$=8.9 \citep{cutri03}
\emph{K2-18 is currently the second brightest M dwarf with a transiting
  habitable zone planet} behind the recently discovered LHS 1140b. Although in the coming years the sample of
habitable zone M dwarf planets is expected to increase dramatically following the launch of \emph{TESS}
\citep{ricker14}, although the majority of TESS planets will be more distant than LHS 1140 \citep{sullivan15}.

Considering the prospect of observational follow-up of K2-18b in transmission spectroscopy, if we
consider an atmosphere that is
cloud-free and dominated by hydrogen, then spectral features from well-mixed near-IR absorbing species
such as water would have amplitudes of $\Delta F/F \sim 10H r_p/R_s^2 \sim 230$ ppm where
$H=k_{\text{B}}T_{\text{eq}}/\mu g$ is the atmospheric scale height, $T_{\text{eq}}$ is the planet's equilibrium
temperature set by the stellar insolation, $\mu$ is the mean molecular weight of the atmosphere, and $g$ is the
surface gravity \citep{millerricci09, kaltenegger09}. If instead the atmosphere is dominated
by heavier elements similar to Earth's (e.g. $\text{N}_2 + \text{O}_2, \mu=29$)
then the transmission signal will be significantly smaller ($\sim 10$ ppm) though potentially
still detectable with JWST with several visits. Because of the brightness of its host star and the low bulk
density of K2-18b, the system offers a unique opportunity to study super-Earth atmospheres receiving
Earth-like insolation in the JWST-era.

\begin{acknowledgements}
  RC is partially supported in this work by the National Science and Engineering Research Council of Canada
  and the Institute for Research on Exoplanets. RC would like to thank Dan Tamayo for his useful discussions
  regarding the dynamical simulations presented in this work and their interpretation. 
  XB, XD, and TF acknowledge the support of CNRS/PNP (Programme national de plan\'etologie).
  D.E. acknowledges support from the National Centre for Competence in Research `PlanetS' of the Swiss National Science
  Foundation (SNSF) and from the European Research Council (ERC) under the European Union's Horizon 2020 research and
  innovation programme (grant agreement No 724427).
  NCS acknowledges support from Funda\c{c}\~ao para a Ci\^encia e a Tecnologia (FCT) through national funds
  and from FEDER through COMPETE2020 by the following grants: UID/FIS/04434/2013 \&
  POCI--01--0145-FEDER--007672 and PTDC/FIS-AST/1526/2014 \& POCI--01--0145-FEDER--016886. NCS also
  acknowledges the support from FCT through Investigador FCT contract IF/00169/2012/CP0150/CT0002.
  Some of the data presented in this paper were obtained from the
  Mikulski Archive for Space Telescopes
  (MAST). STScI is operated by the Association of Universities for Research in Astronomy, Inc., under
  NASA contract NAS5-26555. Support for MAST for non-HST data is provided by the NASA Office of Space
  Science via grant NNX09AF08G and by other grants and contracts.
\end{acknowledgements}

\bibliographystyle{aa}
\bibliography{refs}

\newgeometry{margin=1cm} 
\begin{landscape}
\begin{table*}
\tiny
\renewcommand{\arraystretch}{0.7}
\caption[]{Model parameters}
\label{table:k2184}
\begin{tabular}{lcccc}
\hline \\ [-1ex]
Parameter & \multicolumn{4}{c}{Maximum a-posteriori values with $16^{\text{th}}$ and $84^{\text{th}}$ percentiles} \smallskip\\
\hline \\ [-1ex]
\emph{Stellar Parameters} & & & & \smallskip \\
2MASS Photometry & \multicolumn{4}{c}{$J$=9.763$\pm$0.028, $H$=9.135$\pm$0.026, $K_s$=8.899$\pm$0.019} \\
Stellar mass, $M_s$ [M$_{\odot}$]  &  \multicolumn{4}{c}{0.359 $\pm$ 0.047}   \\
Stellar radius, $R_s$ [R$_{\odot}$]   &  \multicolumn{4}{c}{0.411 $\pm$ 0.038}   \\
Effective temperature, $T_{\text{eff}}$ [K]  & \multicolumn{4}{c}{3457 $\pm$ 39} \\
Distance, $d$ [pc] & \multicolumn{4}{c}{$34 \pm 4$} \\
Rotation period, $P_{\text{rot}}$ [days] & \multicolumn{4}{c}{$38.6^{+0.6}_{-0.4}$} \\
Systemic velocity, $\gamma_0$ [m s$^{-1}$] & \multicolumn{4}{c}{$653.7 \pm 0.9^{\bullet}$} \medskip \\

& \emph{Model 1} & \emph{Model 2} & \emph{Model 3} & \emph{Model 4} \smallskip \\
\emph{GP hyperparameters} & & & & \smallskip \\
Covariance amplitude, $a$ [\mps{]} & $0.1^{+2.8}_{-0.1}$ & N.S.$^{\circ}$ & N.S. & $-$ \\
Exponential timescale, $\lambda$ [days] &  $59.1^{+19.1}_{-11.2}$ & N.S. & N.S. & $-$\\
Coherence, $\Gamma$ & $1.2^{+0.6}_{-0.3}$ & N.S. & N.S. & $-$ \\
Periodic timescale, $P_{\text{GP}}$ [days] & $38.6^{+1.1}_{-0.4}$ & N.S. & N.S. & $-$ \\
Additive jitter, $s$ [\mps{]} & $0.25^{+0.97}_{-0.06}$ & $0.45^{+0.85}_{-0.23}$ & $0.22^{+1.09}_{-0.04}$ & $-$ \medskip \\

\emph{K2-18c} & & & & \smallskip \\
Period, $P_c$ [days] & $8.962 \pm 0.008$ & $8.964 \pm 0.010$ & $8.965 \pm 0.010$ & $8.966^{+0.005}_{-0.010}$ \\
Time of inferior conjunction, $T_{0,c}$ [BJD-2,450,000] & 7264.55 $\pm$ 0.46 & $7264.55 \pm 0.51$ & $7264.49^{+0.58}_{-0.45}$ & $7264.48^{+0.61}_{-0.31}$ \\
Radial velocity semi-amplitude, $K_c$ [\mps{]} & $4.63 \pm 0.72$ & $4.74^{+0.71}_{-0.98}$ & $4.52 \pm 0.82$ & $4.63^{+0.82}_{-0.58}$  \\
$h_c =\sqrt{e_c}\cos{\omega_c}$ & $-0.37^{+0.41}_{-0.13}$ & $-0.46^{+0.45}_{-0.09}$ & $-0.39^{+0.51}_{-0.10}$ & $-0.51^{+0.57}_{-0.05}$ \\
$k_c =\sqrt{e_c}\sin{\omega_c}$ & $0.05^{+0.28}_{-0.21}$ & $-0.04^{+0.35}_{-0.13}$ & $0.00^{+0.35}_{-0.17}$ & $0.01^{+0.32}_{-0.17}$ \medskip \\
Semi-major axis, $a_c$ [AU] & $0.060 \pm 0.003$  & $0.060 \pm 0.003$ & $0.060 \pm 0.003$ & $0.060 \pm 0.003$ \\
Eccentricity, $e_c$   & $< 0.47^{\ast}$ & - & - & - \\
Minimum planet mass, $m_{p,c} \sin{i_c}$ [M$_{\oplus}$] & $7.51 \pm 1.33$ & $7.68^{+1.33}_{-1.72}$ & $7.33 \pm 1.48$ & $7.51^{+1.48}_{-1.15}$  \\
Equilibrium temperature, $T_{\text{eq},c}$ [K] & \\
\hspace{2pt} Bond albedo of 0.3 & 363 $\pm$ 14 & 363 $\pm$ 14 & 363 $\pm$ 14 & 363 $\pm$ 14  \medskip \\

\emph{K2-18b} & & & \smallskip \\
Period, $P_b$ [days] & $32.93963 \pm 1.0 \times 10^{-4}$ & $32.93962 \pm 1.1 \times 10^{-4}$ & $32.93961 \pm 1.0 \times 10^{-4}$ & $32.93960 \pm 9.3 \times 10^{-5}$ \\
Time of inferior conjunction, $T_{0,b}$ [BJD-2,450,000] & $7264.39157 \pm 5.9 \times 10^{-4}$ & $7264.39133 \pm 6.4 \times 10^{-4}$ & $7264.39155^{+0.0006}_{-0.0008}$ & $7264.39135^{+0.0007}_{-0.0005}$ \\
Radial velocity semi-amplitude, $K_b$ [\mps{]} & $3.18 \pm 0.71$ & $3.29^{+0.71}_{-0.64}$ & $3.26^{+0.63}_{-0.85}$ & $3.25^{+0.60}_{-0.41}$  \\
$h_b =\sqrt{e_b}\cos{\omega_b}$ & $0.33^{+0.08}_{-0.28}$ & $0.33^{+0.06}_{-0.30}$ & $0.31^{+0.07}_{-0.37}$ & $0.32^{+0.06}_{-0.27}$ \\
$k_b =\sqrt{e_b}\sin{\omega_b}$ & $-0.16^{+0.42}_{-0.21}$ & $-0.16^{+0.42}_{-0.19}$ & $-0.10^{+0.41}_{-0.25}$ & $-0.11^{+0.31}_{-0.22}$ \medskip \\
Semi-major axis, $a_b$ [AU] & $0.143 \pm 0.006$  & $0.143 \pm 0.006$ & $0.143 \pm 0.006$ & $0.143 \pm 0.006$ \\
Eccentricity, $e_b$   & $< 0.43^{\ast}$ & - & - & - \\
Planet mass, $m_{p,b}$ [M$_{\oplus}$]$^{\dagger}$ & $7.96 \pm 1.91$ & $8.23^{+1.92}_{-1.76}$ & $8.16^{+1.73}_{-2.24}$ & $8.13^{+1.66}_{-1.25}$  \\
Planet density, $\rho_{p,b}$ [$\mathrm{g\;cm^{-3}}$]$^{\ddagger}$ & $3.7 \pm 0.9$ & $3.8^{+0.9}_{-0.8}$ & $3.8^{+0.8}_{-1.1}$ & $3.8^{+0.8}_{-0.6}$ \\
Surface gravity, $g$ [$\mathrm{m\;s^{-2}}$]$^{\ddagger}$  & $15.2 \pm 3.7$ & $15.7^{+3.7}_{-3.4}$ & $15.6^{+3.3}_{-4.3}$ & $15.6^{+3.2}_{-2.4}$  \\
Escape velocity, $v_{\text{esc},b}$ [$\mathrm{km\;s^{-1}}$]$^{\ddagger}$ & $21.0 \pm 2.5$ & $21.4^{+2.5}_{-2.3}$ & $21.3^{+2.3}_{-2.9}$ & $21.2^{+2.2}_{-1.6}$  \\
Equilibrium temperature, $T_{\text{eq},b}$ [K] & \\
\hspace{2pt} Bond albedo of 0.3 & 235 $\pm$ 9 & 235 $\pm$ 9 & 235 $\pm$ 9 & 235 $\pm$ 9 \medskip \\

\emph{Model diagnostics} & & & \smallskip \\
Median $\ln{\mathcal{L}}$ from Cross-Validation & $-1.462 \pm 0.013$ & $-1.574 \pm 0.017$ & $-1.600 \pm 0.018$ & $-1.548 \pm 0.017$ \\
$\ln{\mathcal{L}_1} - \ln{\mathcal{L}_i}$ & 0 & $0.112 \pm 0.021$ & $0.138 \pm 0.022$ & $0.086 \pm 0.021$ \\

\hline
\end{tabular}
\begin{list}{}{}
\item {\bf{Notes.}} $^{(\bullet)}$ from the Model 1 RV modeling. \\
  $^{(\circ)}$ N.S. stands for `no solution' and occurs when the parameter's marginalized posterior 
  PDF is unconstrained by the data. \\
  $^{(\ast)}$ upper limit based on the 99\% confidence interval from the RV data analysis and 
  conditioned on the dynamical stability constraints from Sect.~\ref{sect:dynam}. \\
  $^{(\dagger)}$ assuming the measured orbital inclination of K2-18b from \citetalias{benneke17};
  $i_b = 89.5785^{+0.0079}_{-0.0088}$ degrees. \\
  $^{(\ddagger)}$ assuming the measured radius of K2-18b from \citetalias{benneke17};
  $r_p = 2.279 \pm 0.026$ R$_{\oplus}$. \\
  M$_{\odot} = 1.988499 \times 10^{30}$ kg, R$_{\odot} = 6.955 \times 10^{8}$ m, M$_{\oplus} = 6.045898 \times 10^{24}$ kg, R$_{\oplus} = 6.378137 \times 10^{6}$ m.
\end{list}
\end{table*}
\end{landscape}

\begin{table}
\tiny
\renewcommand{\arraystretch}{0.7}
\centering
\caption[]{HARPS Time-Series}
\label{table:data}
\begin{tabular}{ccccccc}
\hline \\ [-1ex]
BJD-2,450,000 & RV & $\sigma$RV & S index & H$\alpha$ & FWHM & BIS \\
& [m s$^{-1}$] & [m s$^{-1}$] &  &  & & \\
\hline \\ [-1ex]
7117.565870 & 658.15 & 4.30 & 0.551 & 0.06337 & 3.050 & 9.397 \\
7146.526948 & 654.90 & 2.94 & 0.851 & 0.06646 & 3.068 & -3.008 \\
7146.646070 & 660.66 & 3.98 & 0.537 & 0.06850 & 3.079 & -6.150 \\
7148.518851 & 649.21 & 4.62 & 0.973 & 0.06667 & 3.071 & 8.583 \\
7199.503915 & 656.11 & 3.43 & 0.558 & 0.06597 & 3.090 & 16.922 \\
7200.503114 & 656.48 & 2.79 & 0.290 & 0.06625 & 3.080 & 19.343 \\
7204.491167 & 648.52 & 4.38 & 0.368 & 0.06409 & 3.076 & -10.068 \\
7390.845075 & 655.54 & 2.65 & 1.040 & 0.06721 & 3.106 & -0.196 \\
7401.779223 & 649.19 & 2.66 & 0.960 & 0.06640 & 3.105 & 1.064 \\
7403.826871 & 651.38 & 2.87 & 1.287 & 0.06598 & 3.106 & 5.332 \\
7404.814521 & 654.49 & 3.78 & 1.410 & 0.06679 & 3.095 & 6.633 \\
7405.789149 & 655.54 & 2.71 & 1.066 & 0.06644 & 3.106 & 7.934 \\
7407.773473 & 652.66 & 4.36 & - & 0.06680 & 3.048 & 0.311 \\
7410.791609 & 651.99 & 4.35 & 1.296 & 0.07166 & 3.114 & -2.344 \\
7412.810195 & 661.94 & 3.50 & 0.677 & 0.06618 & 3.094 & -8.771 \\
7417.787334 & 649.38 & 3.41 & 0.724 & 0.06615 & 3.092 & 7.447 \\
7418.799229 & 648.38 & 3.67 & 1.402 & 0.06842 & 3.100 & 2.274 \\
7420.791577 & 658.68 & 2.54 & 1.170 & 0.06738 & 3.099 & 11.285 \\
7421.794046 & 658.62 & 2.30 & 1.509 & 0.07463 & 3.099 & 11.310 \\
7422.781258 & 657.47 & 2.66 & 0.986 & 0.06505 & 3.104 & 7.032 \\
7424.777426 & 658.79 & 2.86 & 0.924 & 0.06642 & 3.104 & 5.314 \\
7425.850669 & 652.07 & 2.81 & 0.871 & 0.06636 & 3.088 & -0.471 \\
7446.704487 & 652.60 & 2.67 & 0.933 & 0.06481 & 3.108 & 1.511 \\
7447.830725 & 659.17 & 3.87 & 0.862 & 0.06539 & 3.085 & 0.282 \\
7448.686909 & 656.59 & 3.26 & 0.713 & 0.06470 & 3.098 & 0.282 \\
7450.675147 & 660.91 & 3.33 & 0.917 & 0.06521 & 3.102 & -8.218 \\
7451.677499 & 655.48 & 2.70 & 1.063 & 0.06575 & 3.111 & -5.083 \\
7452.695705 & 663.48 & 2.75 & 1.112 & 0.06532 & 3.093 & 1.667 \\
7453.701988 & 655.18 & 2.38 & 1.038 & 0.06689 & 3.104 & 10.356 \\
7456.704230 & 658.87 & 3.31 & 0.887 & 0.06586 & 3.111 & -17.835 \\
7457.683261 & 658.28 & 3.81 & 1.010 & 0.06615 & 3.101 & 12.792 \\
7458.660021 & 656.62 & 3.47 & 1.404 & 0.06634 & 3.099 & 9.160 \\
7472.784787 & 641.61 & 3.12 & 1.058 & 0.06653 & 3.094 & -1.001 \\
7473.684129 & 648.26 & 3.08 & 0.908 & 0.06529 & 3.096 & -1.293 \\
7474.737446 & 645.84 & 3.90 & 0.783 & 0.06621 & 3.085 & -6.991 \\
7475.698658 & 655.63 & 3.94 & 0.331 & 0.06647 & 3.094 & -0.455 \\
7476.707703 & 652.59 & 3.26 & 0.983 & 0.06517 & 3.091 & -0.985 \\
7477.674398 & 657.56 & 3.37 & 0.816 & 0.06554 & 3.100 & -4.990 \\
7478.631994 & 652.31 & 4.64 & 0.682 & 0.06713 & 3.093 & -9.857 \\
7479.737617 & 649.94 & 3.17 & 1.116 & 0.06541 & 3.086 & -9.996 \\
7486.661319 & 666.10 & 3.24 & 1.168 & 0.06527 & 3.096 & -10.681 \\
7487.617699 & 667.97 & 4.55 & 0.573 & 0.06504 & 3.091 & -0.986 \\
7488.670507 & 656.68 & 3.56 & 0.922 & 0.06431 & 3.086 & -8.201 \\
7567.516862 & 659.76 & 4.00 & 0.993 & 0.06530 & 3.096 & 10.014 \\
7576.473152 & 657.32 & 4.27 & 0.989 & 0.06587 & 3.115 & 7.131 \\
7584.477527 & 657.94 & 5.55 & 0.725 & 0.06563 & 3.080 & 16.445 \\
7786.842858 & 645.83 & 3.49 & 1.146 & 0.06430 & 3.093 & -1.668 \\
7787.825672 & 658.25 & 4.66 & 0.944 & 0.06581 & 3.092 & 5.481 \\
7790.828228 & 655.56 & 3.49 & 1.092 & 0.06500 & 3.088 & 1.878 \\
7791.843445 & 653.87 & 3.70 & 0.700 & 0.06414 & 3.093 & -3.719 \\
7792.815105 & 653.38 & 3.47 & 0.893 & 0.06861 & 3.083 & 6.408 \\
7801.827514 & 652.57 & 3.18 & 0.812 & 0.06518 & 3.094 & 12.389 \\
7802.790293 & 648.98 & 3.14 & 0.860 & 0.06504 & 3.094 & -3.525 \\
7803.809311 & 649.61 & 3.27 & 0.926 & 0.06616 & 3.091 & 18.332 \\
7810.806284 & 654.41 & 4.71 & 0.905 & 0.06593 & 3.098 & 6.496 \\
7814.760772 & 652.53 & 4.35 & - & 0.06632 & 3.098 & -0.563 \\
7815.759421 & 656.22 & 5.06 & - & 0.06536 & 3.086 & 15.990 \\
7817.748614 & 665.63 & 2.89 & 0.998 & 0.06423 & 3.103 & 7.162 \\
7830.668729 & 647.15 & 3.18 & 0.866 & 0.06476 & 3.084 & -1.992 \\
7832.659387 & 647.89 & 4.31 & 1.029 & 0.06292 & 3.079 & 14.048 \\
7834.636450 & 656.64 & 4.31 & 1.025 & 0.06407 & 3.091 & 9.870 \\
7835.596293 & 654.97 & 4.36 & 1.004 & 0.06308 & 3.093 & 11.521 \\
7836.626075 & 656.46 & 4.71 & 0.792 & 0.06491 & 3.093 & -2.004 \\
7839.650934 & 644.07 & 3.74 & 0.774 & 0.06437 & 3.076 & 5.253 \\
7841.638147 & 648.49 & 3.92 & 0.810 & 0.06487 & 3.089 & 0.543 \\
7843.648062 & 654.65 & 3.92 & 0.408 & 0.06404 & 3.090 & 12.825 \\
7844.626814 & 659.07 & 3.85 & 0.722 & 0.06417 & 3.087 & 1.584 \\
7846.692642 & 655.96 & 3.75 & 0.923 & 0.06435 & 3.086 & 5.210 \\
7847.693240 & 651.46 & 3.77 & 0.903 & 0.06363 & 3.094 & 8.538 \\
7848.677241 & 655.24 & 3.82 & 0.848 & 0.06445 & 3.102 & 1.280 \\
7849.696944 & 653.92 & 3.34 & 0.853 & 0.06440 & 3.082 & 9.616 \\
7872.656844 & 653.72 & 3.92 & 0.913 & 0.06371 & 3.082 & -5.325 \\
7873.525484 & 652.07 & 4.95 & 0.813 & 0.06478 & 3.094 & 13.489 \\
7874.671695 & 650.86 & 4.80 & 0.886 & 0.06543 & 3.088 & 1.857 \\
7875.596914 & 643.77 & 3.69 & 0.913 & 0.06467 & 3.077 & 5.739 \\
\hline
\end{tabular}
\end{table}

\end{document}